\documentclass{article}

\usepackage{arxiv}

\usepackage[utf8]{inputenc} 
\usepackage[T1]{fontenc}    
\usepackage{hyperref}       
\usepackage{url}            
\usepackage{booktabs}       
\usepackage{amsfonts}       
\usepackage{nicefrac}       
\usepackage{microtype}      
\usepackage{graphicx}
\usepackage{graphicx}%
\usepackage{multirow}%
\usepackage{amsmath,amssymb,amsfonts}%
\usepackage{amsthm}%
\usepackage{mathrsfs}%
\usepackage[title]{appendix}%
\usepackage{xcolor}%
\usepackage{textcomp}%
\usepackage{manyfoot}%
\usepackage{algorithm}%
\usepackage{algorithmicx}%
\usepackage{algpseudocode}%
\usepackage{listings}%
\usepackage{float}
\usepackage[T1]{fontenc}
\usepackage{bm}
\usepackage{adjustbox}
\usepackage{subfig}
\title{Sparse Functional Data Classification via Bayesian Aggregation}

\author{
Ahmad Talafha \\
  Department of Mathematics\\
St. Edward's University\\
3001 S Congress Ave\\
 Austin, TX 78704 \\
  \texttt{atalafha@stedwards.edu} 
  }

\begin{document}
\maketitle
\begin{abstract}
Sparse functional data frequently arise in real-world applications, posing significant challenges for accurate classification. To address this, we propose a novel classification method that integrates functional principal component analysis (FPCA) with Bayesian aggregation. Unlike traditional ensemble methods, our approach combines predicted probabilities across bootstrap replicas and refines them through Bayesian calibration using Bayesian generalized linear models (Bayesian GLMs). We evaluated the performance of the proposed method against single classifiers and conventional ensemble techniques. The simulation results demonstrate that Bayesian aggregation improves the classification accuracy over conventional methods. Finally, we validate the approach through three real-data analyses.
\end{abstract}


\section{Introduction}\label{sec1}

The advancement of measurement technologies has enabled the collection of data in the form of curves or functions across various fields, such as meteorology \cite{Jiang2020Principal, tadic2019example, zuska2019application, hael2021modeling}, health sciences \cite{froslie2013shape, sanchez2014functional, yang2024predicting, burns2013functional, che2017trajectory, salvatore2016exploring, karuppusami2022functional, woo2023exploring}, and genomics \cite{miao2020increased, Yunqing2020fun}. These data, often referred to as functional or longitudinal data, are inherently infinite-dimensional since they are represented as continuous functions rather than discrete points. Functional data analysis (FDA) has emerged as a powerful framework for analyzing such data, with dimensionality reduction being a critical step due to the high-dimensional nature of functional observations. Among the various dimensionality reduction techniques, functional principal component analysis (FPCA) has gained significant attention due to its ability to capture the primary modes of variation in functional data through data-driven bases known as functional principal component (FPC) scores (\cite{silverman1996smoothed}).

In many real-world applications, functional data are observed sparsely or irregularly, meaning that each curve is measured at only a few time points, and these points may vary across individuals. This sparsity poses significant challenges for traditional FDA methods, as estimating the covariance structure and FPC scores becomes computationally intensive and prone to bias. To address these challenges, \cite{james2000principal} proposed a reduced-rank mixed-effects model for sparse functional data, which estimates FPC scores using an Expectation-Maximization (EM) algorithm. More recently, \cite{yao2005functional} introduced the principal analysis by conditional expectation (PACE) method, which provides unbiased estimates of FPC scores for sparse functional data by leveraging conditional expectation techniques.

Although FPCA has been widely used for functional data classification tasks, several alternative methods have been proposed. For instance, \cite{james2001functional} developed a functional linear discriminant analysis (FLDA) for sparse functional data using an EM algorithm. \cite{james2002generalized} and \cite{muller2005generalized} extended the generalized linear model to functional data, while \cite{leng2006classification} applied functional logistic regression based on FPC scores to temporal gene expression data. Other approaches include support vector machines (SVM) for functional data \cite{lee2010functional, rossi2006support} and kernel-induced random forests for functional data classification (\cite{fan2015functional}). Despite these advancements, the classification of sparse functional data remains a challenging task, particularly when the data are irregularly sampled or contain measurement noise.

Bayesian methods have gained popularity in functional data analysis due to their ability to handle uncertainty and incorporate prior knowledge (\cite{gelman2013bayesian}). For example, Bayesian generalized linear models (Bayesian GLM) have been widely used for model calibration and prediction (\cite{Neal1996}). In the context of ensemble learning, Bayesian model averaging (\cite{Raftery1997}) and Bayesian additive regression trees (BART) (\cite{Chipman2010}) have been shown to improve classification accuracy by combining predictions from multiple weak learners. These methods are particularly relevant to our proposed Bayesian aggregation approach, which aggregates predicted probabilities across bootstrap replicas and refines them using Bayesian calibration.

Unlike previous studies, we propose a novel classification method for sparse functional data that integrates FPCA with Bayesian aggregation. While traditional ensemble methods such as bagging enhance predictions by combining classifiers trained on bootstrap samples (\cite{breiman1996bagging}), our approach extends this framework by incorporating Bayesian calibration to refine the predicted probabilities. Specifically, instead of aggregating hard classification labels or uncalibrated probability estimates, we first compute predicted probabilities from classifiers trained on bootstrap samples. These probabilities are then aggregated and refined using a Bayesian generalized linear model (Bayesian GLM), which explicitly accounts for uncertainty in the predictions and adjusts for potential miscalibrations. This Bayesian calibration step ensures that the final probability estimates are both accurate and well-calibrated, leading to more reliable classifications.

Previous work by \cite{bauer1999empirical} compared bagging, AdaBoost, and other ensemble techniques on decision trees and naive Bayes classifiers using large-scale datasets, demonstrating the effectiveness of ensemble methods in improving classification performance. Similarly, \cite{kim2022Bootstrap} developed a classification method for sparse functional data using FPCA and bagging, demonstrating improved performance over single classifiers through simulations and real-data analyses, while \cite{Shinde01052014} applied bagging to kernel principal component analysis to enhance preimage estimation. However, these methods often rely on majority voting or weighted averaging, which do not explicitly model the uncertainty in classifier outputs. By contrast, our Bayesian aggregation approach leverages the strengths of Bayesian inference to handle uncertainty and improve robustness, particularly in the context of sparse and irregularly sampled functional data.

By extending these ideas to functional data, we develop a Bayesian aggregation model for sparse functional data that combines classifiers based on FPC scores derived from bootstrap samples. The key advancement lies in the Bayesian calibration step, which adjusts the aggregated probabilities using a Bayesian GLM. This step incorporates prior knowledge and updates the probability estimates based on the observed data, resulting in better-calibrated predictions. Our method improves classification accuracy and provides a principled way to quantify uncertainty, making it particularly suitable for applications where data sparsity and measurement noise are significant challenges.

Our contribution to the literature lies in enhancing classification performance for sparse functional data and introducing a Bayesian ensemble technique to functional principal components-based classification models. This approach bridges the gap between traditional ensemble methods and Bayesian inference, offering a robust and interpretable framework for classifying sparse functional data.

The remainder of the paper is organized as follows: Section 2 provides a comprehensive review of the FPCA method for sparse functional data, with a particular focus on the PACE approach. It also covers classification based on FPC scores, introduces the proposed Bayesian aggregation method, and describes its implementation in detail. Section 3 presents the results of our simulation studies, evaluating the performance of the proposed method under various settings. It also applies the method to real-world datasets, demonstrating its practical effectiveness. Finally, Section 4 concludes the paper with a discussion of key findings and potential extensions for future research.

\section{Methods} \label{methods}

\subsection{FPCA for Sparse Functional Data}
\label{subsection:fpca}

Functional principal component analysis (FPCA) is a powerful tool for dimensionality reduction in functional data analysis. It is based on the Karhunen-Lo\`eve representation, which decomposes a random function into a series of orthogonal basis functions. Let \( X(t) \) for \( t \in T \) denote a square-integrable random process in \( L^2(T) \), with mean function
\begin{equation*}
E[X(t)] = \mu(t)
\end{equation*}
and covariance function
\begin{equation*}
G(s,t) = \text{cov}[X(s),X(t)] \quad \text{for } s,t \in T.
\end{equation*}

According to Mercer's theorem (\cite{INDRITZ_1963}), the covariance function can be expressed as:
\begin{equation*}
G(s,t) = \sum_{k=1}^{\infty} \lambda_k \varphi_k(s) \varphi_k(t), \quad \text{for } s,t \in T.
\end{equation*}
where \( \lambda_1 \geq \lambda_2 \geq \cdots \geq 0 \) are nonnegative eigenvalues satisfying \( \sum\limits_{k=1}^{\infty} \lambda_k < \infty \), and \( \varphi_k \) are the corresponding orthonormal eigenfunctions. Given \( n \) random curves,
\begin{equation*}
{X} = [X_1(t), \dots, X_n(t)],
\end{equation*}
the Karhunen-Lo\`eve expansion \cite{karhunen_1947, loeve_1948} of \( X_i(t) \) is given by:
\begin{equation*}
X_i(t) = \mu(t) + \sum_{k=1}^{\infty} \xi_{ik} \varphi_k(t), \quad t \in T,
\end{equation*}
where
\begin{equation*}
\xi_{ik} = \int_T (X_i(t) - \mu(t)) \varphi_k(t) \, dt
\end{equation*}
are uncorrelated random variables with mean 0 and variance \( \lambda_k \). In practice, this infinite series is truncated to a finite number of terms, yielding the approximation:
\begin{equation*}
X_i(t) \approx \mu(t) + \sum_{k=1}^{K} \xi_{ik} \varphi_k(t), \quad t \in T,
\end{equation*}
where \( K \) is the number of basis functions. The choice of \( K \) is typically guided by the proportion of variance explained (PVE), although alternative criteria such as the Akaike information criterion (AIC) or Bayesian information criterion (BIC) can also be used \cite{Li2013,yao2005functional}.

While FPCA is well-suited for densely observed functional data, its application to sparse or irregularly sampled data poses significant challenges. In sparse settings, each curve \( X_i(t) \) is observed at only a few time points, making it difficult to directly compute the covariance function \( G(s,t) \) and estimate the functional principal component (FPC) scores \( \xi_{ik} \). Naively applying conventional FPCA to sparse data can lead to biased estimates and poor performance.

To address these challenges, \cite{james2000principal} proposed a reduced-rank mixed-effects model for sparse functional data. More recently, \cite{yao2005functional} introduced the principal analysis by conditional expectation (PACE) method, which provides unbiased estimates of FPC scores for sparse functional data.

Consider the \( i \)-th curve
\begin{equation}
\bm{X}_i = (X_i(t_{i1}), \dots, X_i(t_{in_i}))^{\top},
\end{equation}
where \( t_{ij} \in T \) is the \( j \)-th time point observed for the \( i \)-th curve, and \( n_i \) is the number of observations for the \( i \)-th curve. Let \( \bm{Z}_i = (Z_i(t_{i1}), \dots, Z_i(t_{in_i}))^{\top}\) denote the observed curve, which includes measurement errors \( \bm{\varepsilon}_i=(\varepsilon_i(t_{i1}), \dots, \varepsilon_i(t_{in_i}))^{\top}\), given by:
\begin{equation}
\bm{Z}_i = \bm{X}_i + \bm{\varepsilon}_i.
\end{equation}
The relationship between the observed and true curves is given by:
\begin{equation}
Z_i(t_{ij}) = X_i(t_{ij}) + \varepsilon_i(t_{ij}) = \mu(t_{ij}) + \sum_{k=1}^{\infty} \xi_{ik} \varphi_k(t_{ij}) + \varepsilon_i(t_{ij}), \quad t_{ij} \in T,
\end{equation}
where \( \varepsilon_i(t_{ij}) \) are independent and identically distributed (i.i.d.) errors with mean $0$ and variance \( \sigma^2 \), and are assumed to be independent of the FPC scores \( \xi_{ik} \).

Under the assumption that \( \xi_{ik} \) and \( {\varepsilon}_i \) are jointly Gaussian, the best linear unbiased prediction (BLUP) of \( \xi_{ik} \) is given by:
\begin{equation}
\hat{\xi}_{ik} = E[\xi_{ik} \mid \bm{Z}_i] = \lambda_k \bm{\varphi}_k^{\top} \bm{\Sigma}_{\bm{Z}_i}^{-1} (\bm{Z}_i - \bm{\mu}_i),
\end{equation}
where \( \bm{\varphi}_{ik} = (\varphi_k(t_{i1}), \dots, \varphi_k(t_{in_i}))^{\top} \) is the \( i \)-th FPC function, \(\bm{\mu}_i=(\mu_i(t_{i1}), \dots, \mu_i(t_{in_i}))^{\top}\) and
\begin{equation}
\bm{\Sigma}_{Z_i} = \text{cov}(\bm{Z}_i, \bm{Z}_i) = \text{cov}(\bm{X}_i, \bm{X}_i) + \sigma^2 \bm{I}_{n_i}.
\end{equation}

\subsection{Classification with FPC Scores} \label{subsec:class_fpc}

Classification models based on the \( K \) functional principal component (FPC) scores, \( \hat{\xi}_{ik} \), for \( k = 1, \ldots, K \), derived in Section \ref{subsection:fpca}, provide a versatile framework for analyzing functional data. We employ several classifiers, including logistic regression (logit), gradient boosting machines (GBM), random forests (RF), linear discriminant analysis (LDA), quadratic discriminant analysis (QDA), and naive Bayes, to classify observations using the FPC scores.

Logistic regression is implemented using the functional generalized linear model (FGLM) \cite{james2002generalized, muller2005generalized}. The decision function is expressed as:
\[
    h(Z_i) = \alpha + \sum_{k=1}^{K} \beta_k \xi_{ik},
\]
where \( \alpha \) and \( \beta_k \) are coefficients, and \( \xi_{ik} \) are the FPC scores. A logit link function maps \( h(Z_i) \) to the probability space, and the predicted class is determined by a threshold of 0.5.

GBM and RF are ensemble methods that utilize the FPC scores. GBM builds an additive model by optimizing a differentiable loss function (\cite{Friedman2001GreedyFA}), while RF constructs multiple decision trees and outputs the mode of the classes (\cite{breiman2001random}). Both methods are robust to overfitting and capture complex interactions in the data.

LDA assumes that the FPC scores within each class follow a multivariate normal distribution with a common covariance matrix (\cite{fisher1936use}). The discriminant score for class \( j \) is:
\[
    \delta_j(\boldsymbol{\xi}_i) = \boldsymbol{\xi}_i^\top \mathbf{\Sigma}^{-1} \boldsymbol{\mu}_j - \frac{1}{2} \boldsymbol{\mu}_j^\top \mathbf{\Sigma}^{-1} \boldsymbol{\mu}_j + \log(\pi_j),
\]
where \( \boldsymbol{\mu}_j \) and \( \mathbf{\Sigma} \) are the class mean and common covariance matrix, respectively. QDA relaxes the assumption of equal covariance matrices, allowing class-specific covariance matrices \( \mathbf{\Sigma}_j \) (\cite{hastie2009elements}).

The naive Bayes classifier assumes conditional independence of the FPC scores given the class label (\cite{murphy2012machine}). The conditional probability for class \( j \) is:
\[
    P(Y_i = j \mid \boldsymbol{\xi}_i) \propto \prod_{k=1}^{K} P(\xi_{ik} \mid Y_i = j) \pi_j,
\]
where \( P(\xi_{ik} \mid Y_i = j) \) is modeled as a univariate Gaussian.

While support vector machines (SVM) (\cite{cortes1995support}), both linear and radial basis function (RBF) kernels, are commonly used for classification, they were not included in this analysis. Although SVMs can be applied to the FPC scores, their exclusion is motivated by their lack of inherent probabilistic outputs. Unlike models introduced here, which naturally produce probability estimates, SVMs require additional calibration techniques such as Platt scaling (\cite{platt1999probabilistic}). This additional calibration introduces computational complexity and potential bias. Given the emphasis on probabilistic classification for Bayesian aggregation, SVMs were omitted in favor of models that naturally yield class probability estimates.

For the RF model, we select the tuning parameters through an iterative search process that optimizes the number of randomly selected predictors at each split. This is done using a stepwise adjustment factor, aiming to improve model accuracy while maintaining computational efficiency. The optimal parameters are chosen based on minimizing the out-of-bag (OOB) error. For the GBM model, we perform hyperparameter tuning by evaluating different combinations of the number of trees, interaction depth, learning rate (shrinkage), and the minimum number of observations per terminal node. We select the best tuning parameters through a 5-fold cross-validation of the entire training set, ensuring that the model generalizes well and avoids overfitting.

These classifiers represent single-model approaches to functional data classification, each leveraging FPC scores as covariates to make predictions. While these methods provide valuable insights, their performance can vary depending on the structure of the data, the degree of sparsity, and the inherent complexity of the classification task. To enhance predictive accuracy and model stability, ensemble strategies are employed. The next subsection explores different ensemble techniques, including majority voting, out-of-bag (OOB) error weighting, and Bayesian aggregation, which collectively aim to refine classification decisions by leveraging multiple classifiers and improving probability calibration.

\subsection{Ensemble and Bayesian Aggregation}
\label{AggregationMethods}

In ensemble learning, combining predictions from multiple classifiers can enhance overall model performance and robustness. Aggregation methods aim to consolidate these predictions into a single decision rule, leveraging the strengths of individual models while mitigating their weaknesses. Below, we describe two widely used aggregation techniques: majority vote and out-of-bag (OOB) error weighted vote. These methods serve as benchmarks for evaluating the performance of our proposed Bayesian aggregation framework.

\subsubsection{Majority vote}
\label{Majorityvote}

The majority vote method is one of the simplest and most commonly used aggregation techniques. It selects the class that receives the highest number of votes from all \( B \) classifiers. Let \( \mathcal{C} = \{ c_1, c_2, \dots, c_M \} \) denote a set of \( M \) classifiers applied to the functional principal component (FPC) scores. Formally, the bagged classifier is defined as:

\begin{equation} \label{eq:bag_mv}
   \hat{y}_{\text{bag}} = \arg \max_{j \in \{0,1\}} \frac{1}{B} \sum_{b=1}^{B} \mathbb{I} \{ \hat{c}^{(b)}(x) = j \} 
\end{equation}

where \( \mathbb{I}(\cdot) \) is the indicator function, which equals 1 if the condition is true and 0 otherwise. While majority voting is straightforward and computationally efficient, it does not account for the varying confidence levels or performance of individual classifiers, potentially leading to suboptimal results when classifiers exhibit significant differences in reliability.

\subsubsection{OOB error weighted vote}
\label{OOBweightedvote}

The out-of-bag (OOB) error weighted vote is a more refined aggregation method that assigns weights to classifiers based on their performance. The OOB error, a well-known test error estimator for bagging (\cite{breiman1996bagging}), is calculated using the samples not selected during the bootstrapping process. Let \( e_b \) for \( b = 1, \ldots, B \) denote the OOB errors of the bootstrapped models. The weight for each classifier is defined as \( w_b = 1/e_b \), giving higher weights to better-performing models. If \( e_b = 0 \), it is set to the minimum non-zero OOB error to avoid division by zero. The bagged classifier using the OOB error weighted vote is:

\begin{equation} \label{eq:bag_oob}
\hat{y}_{\text{bag}} = \frac{\sum\limits_{b=1}^{B} w_b \hat{c}^{(b)}(x)}{\sum\limits_{b=1}^{B} w_b}, 
\end{equation}

This method improves the majority vote by incorporating the performance of the classifier into the aggregation process, which was shown in \cite{kim2022Bootstrap}. However, it still does not explicitly model uncertainty or account for potential miscalibrations in the classifier outputs, which can limit its effectiveness in practice.
\subsubsection{Bayesian Aggregation Framework}
\label{BayesianAggregation}

To address the limitations of traditional aggregation methods, we propose a Bayesian aggregation framework that refines classifier outputs through Bayesian calibration. Unlike majority voting and out-of-bag (OOB) weighted voting, our approach explicitly models uncertainty in classifier predictions and leverages Bayesian inference to produce well-calibrated probability estimates. The framework consists of two key steps: (1) bootstrap aggregation to reduce variance and (2) Bayesian calibration to correct for biases and miscalibrations.

Let \(\mathcal{D} = \{(\xi_i, y_i)\}_{i=1}^n\) denote the dataset, where \(\xi_i\) represents the functional principal component (FPC) score for observation \(i\), and \(y_i \in \{0, 1\}\) is the binary response variable indicating the true class label. Each classifier \(c_m\) generates an estimated probability \(p_{i,m}\) for the class membership of observation \(i\):

\begin{equation} \label{eq:agg_boot_prob}
  p_{i,m} = P(Y_i = 1 \mid \xi_i, c_m).  
\end{equation}

To enhance the stability and reliability of these probability estimates, we employ a bootstrap resampling strategy. Specifically, each classifier \(c_m\) is trained on \(B\) bootstrap-resampled datasets. For observation \(i\), the aggregated probability estimate from classifier \(c_m\) is computed as the average across all bootstrap replicates:

\begin{equation} \label{eq:aggregated_prob}
    \hat{p}_{i,m} = \frac{1}{B} \sum_{b=1}^{B} p_{i,m}^{(b)},
\end{equation}

where \(p_{i,m}^{(b)}\) represents the probability estimate from the \(b\)-th bootstrap iteration. This aggregation step reduces variability in the classifier outputs, yielding more stable probability estimates.

While bootstrap aggregation reduces variance, it does not explicitly address potential biases or miscalibrations in the classifier outputs. To refine the aggregated probabilities, we introduce a Bayesian calibration step using a Bayesian GLM (\cite{gelman2013bayesian}). For each observation \(i\) with true class label \(y_i\), we assume:

\begin{equation} \label{eq:Bernoulli}
    y_i \mid \pi_{i,m} \sim \text{Bernoulli}(\pi_{i,m}),
\end{equation}

where \(\pi_{i,m}\) represents the calibrated probability of class membership. The calibration is performed using a logistic regression model:

\begin{equation} \label{eq:logit}
\text{logit}(\pi_{i,m}) = \beta_0 + \beta_1 \hat{p}_{i,m}.
\end{equation}

Here, \(\beta_0\) and \(\beta_1\) are regression coefficients estimated within a Bayesian framework. To ensure robustness and prevent overfitting, we impose weakly informative Cauchy prior distributions on the coefficients:

\begin{equation} \label{eq:Cauchy}
\beta_0 \sim \text{Cauchy}(0, \sigma_0^2), \quad \beta_1 \sim \text{Cauchy}(0, \sigma_1^2),
\end{equation}

where \(\sigma_0^2\) and \(\sigma_1^2\) control the degree of regularization for the intercept and slope, respectively. The Cauchy prior is particularly effective in stabilizing estimates in the presence of separation, a common issue in logistic regression.

The posterior distribution for \(\beta_0\) and \(\beta_1\) is estimated using an approximate Expectation-Maximization (EM) algorithm, which adapts the classical iteratively weighted least squares (IWLS) procedure to incorporate the Cauchy priors (\cite{Gelman2008AWI}). After Bayesian calibration, we obtain the refined probability estimates \(\hat{\pi}_{i,m}\) for each observation \(i\) and classifier \(c_m\):

\begin{equation} \label{eq:prob_pi}
    \hat{\pi}_{i,m} = \frac{1}{1 + \exp\left(-(\hat{\beta}_0 + \hat{\beta}_1 \hat{p}_{i,m})\right)}.
\end{equation}

These calibrated probabilities are then used to make classification decisions based on the Bayesian prediction rule:

\begin{equation} \label{eq:bayesian}
   \hat{y}_{\text{bayesian}} = \mathbb{I}\left(\hat{\pi}_{i,m} > 0.5\right), 
\end{equation}

where \(\mathbb{I}(\cdot)\) is the indicator function. The Bayesian aggregation framework thus provides a principled approach to combining classifier outputs while accounting for uncertainty and improving calibration.

\begin{quote}
\textbf{Algorithm 1: Bayesian Aggregated Functional Classifier via Sparse FPCA}
\begin{enumerate}
    \item 
    \begin{enumerate}
        \item For \( b = 1, \dots, B \):
        \begin{enumerate}
            \item Generate a bootstrap resample \(\mathcal{D}^{(b)}\) from the original dataset \(\mathcal{D}\).
            \item Perform FPCA on \(\mathcal{D}^{(b)}\) to extract FPC scores \(\xi_i^{(b)}\) for each observation \(i\), where \(i = 1, \dots, n\).
            \item Estimate the FPC scores using PACE and select \(K\) FPCs.
            \item Train a classifier \(c_m^{(b)}\) using the \(K\) FPC scores \(\xi_i^{(b)}\) and corresponding labels \(y_i^{(b)}\).
        \end{enumerate}
    \end{enumerate}

    \item  \begin{enumerate}
        \item Given a new curve \(Z^*(t)\):
        \begin{enumerate}
            \item For \(b = 1, \dots, B\):
            \begin{enumerate}
                \item Estimate the FPC scores \(\xi^{*(b)}\) for \(Z^*(t)\) using PACE from \(\mathcal{D}^{(b)}\).
                \item Use the \(b\)-th classifier \(c_m^{(b)}\) to obtain the probability estimate \(p^{*(b)}\) using Equation (\ref{eq:agg_boot_prob})
            \end{enumerate}
            \item Compute the aggregated probability estimate for the new curve, $\hat{p}^*$ using Equation (\ref{eq:aggregated_prob})
            \item Perform Bayesian calibration:
            \begin{enumerate}
                \item Fit a Bayesian logistic regression model using Equation (\ref{eq:logit})
                \item Impose weakly informative Cauchy priors on the coefficients using Equation (\ref{eq:Cauchy})
                \item Compute the calibrated probability estimate using Equation (\ref{eq:prob_pi})
            \end{enumerate}
            \item Obtain the final Bayesian aggregated prediction using Equaiton (\ref{eq:bayesian}). 
        \end{enumerate}
    \end{enumerate}
\end{enumerate}
\end{quote}
\subsection{Simulation Study}

We perform simulation studies by referring to the models in \cite{kim2022Bootstrap} to generate functional data in three distinct models. A total of \( n = 200 \) curves are generated, divided equally into two groups labeled \( g \in \{0, 1\} \). Each curve is represented as follows:

\begin{equation}
Z_{gi}(t_{ij}) = \mu_g(t_{ij}) + \sum_{k=1}^3 \xi_{gk} \varphi_k(t_{ij}) + \epsilon_i(t_{ij}), \quad i = 1, \dots, 100, \quad j = 1, \dots, n_{ij},
\end{equation}

where \( g \) denotes the group label, \( \mu_g(t) \) is the mean function of the group \( g \), \( \varphi_k(t) \) are the eigenfunctions, \( \xi_{gk} \) are the scores of the functional principal component (FPC) and \( \varepsilon_i(t) \) is the measurement error. The eigenfunctions \( \varphi_k(t) \) are defined as:

\begin{equation}
\varphi_k(t) = 
\begin{cases}
\cos(\pi k t / 5) / \sqrt{5}, & \text{if } k \text{ is odd}, \\
\sin(\pi k t / 5) / \sqrt{5}, & \text{if } k \text{ is even},
\end{cases}
\end{equation}

for \( k = 1, 2, 3 \). The FPC scores \( \xi_{gk} \) are generated from either a normal distribution or a \( t \)-distribution, depending on the specified distribution type. For the normal distribution, \( \xi_{gk} \sim \mathcal{N}(0, \lambda_{gk}) \), where \( \lambda_{gk} \) are the eigenvalues for group \( g \) and component \( k \). For the \( t \)-distribution, the FPC scores are generated as:

\begin{equation}
\xi_{gk} = t_{gk} \sqrt{\lambda_{gk} / 3}, \quad t_{gk} \sim t_3,
\end{equation}

where \( t_3 \) denotes a \( t \)-distribution with 3 degrees of freedom. The measurement error \( \epsilon_i(t) \) is sampled from \( {N}(0, 0.5^2) \).

We consider three models, each differing in the specification of the mean function \( \mu_g(t) \) and the eigenvalues \( \lambda_{gk} \): (Model A) different mean and variance structure between groups, (Model B): different mean structure but identical variance structure between groups, and (Model C): Identical mean structure, but different variance structure between groups.

The specific forms of \( \mu_g(t) \) and \( \lambda_{gk} \) for each model are provided in Table~\ref{tab:models}.

\begin{table}[h!]
    \caption{Specification of mean functions and eigenvalues for each model.}
    \label{tab:models}
    \centering
    \begin{tabular}{llll}
        \toprule
        Model & \( g \) & \( \mu_g(t) \) &  \( \lambda_{gk} \) \\
        \midrule
        A & 0 & \( t + \sin(t) \) & \( (4, 2, 1) \) \\
          & 1 & \( t + \cos(t) \) & \( (16, 8, 4) \) \\
        \midrule
        B & 0 & \( t + \sin(t) \) & \( (4, 2, 1) \) \\
          & 1 & \( t + \cos(t) \) & \( (4, 2, 1) \) \\
        \midrule
        C & 0 & \( t + \sin(t) \) & \( (4, 2, 1) \) \\
          & 1 & \( t + \sin(t) \) & \( (16, 8, 4) \) \\
        \bottomrule
    \end{tabular}
\end{table}

To simulate sparsity in the functional data, the number of observations \( n_i \) for the \( i \)-th curve is randomly selected from the set \( \{5, 6, \dots, 10\} \). The corresponding observation times \( t_{ij} \) are independently sampled from a uniform distribution over the interval \( [0, 10] \), i.e., \( t_{ij} \sim \text{Uniform}[0, 10] \), and sorted in ascending order.

To introduce outliers, a proportion \( \rho_{\text{out}} \) of the curves are randomly selected, and a fixed shift of \( 5 \) units is added to their observed values. Specifically, for each outlier curve \( i \), the observed values are modified as:

\begin{equation}
Z^{*}_{gi}(t_{ij}) = Z_{gi}(t_{ij}) + 5.
\end{equation}

Additionally, random noise can be added to the observed values. If the noise level \( \sigma^2_{\text{noise}} > 0 \), the observed values are further perturbed as:

\begin{equation}
Z^{**}_{gi}(t_{ij}) = Z^{*}_{gi}(t_{ij}) + \eta_{ij}, \quad \eta_{ij} \sim {N}(0, \sigma^2_{\text{noise}}).
\end{equation}

The simulation study is designed to evaluate classification performance under various data-generating conditions. Each scenario is characterized by a combination of three key factors: the underlying model structure (A, B, or C), the data distribution (Normal or \( t \)--distribution), and two parameters controlling data complexity--\(\rho_{\text{out}}\), which regulates the level of outlier contamination, and \(\sigma^2_{\text{noise}}\), which represents the variance of the added noise. 

Table~\ref{tab:scenarios} summarizes the nine considered scenarios. Also, Table~\ref{tab:scenarios_K} shows a summary statistics of the number of FPC scores ($K$) that are selected in single and bagging models during fitting for each scenario. We select the number of FPC scores, $K$, by a PVE greater than
$0.99$.

\begin{table}[h!]
    \centering
    \caption{Summary of the simulation scenarios used in the study.}
    \label{tab:scenarios}
    \begin{tabular}{cclll}
        \toprule
     Scenario & Model & $\xi_{gk}$ & $\rho_{\text{out}}$ & $\sigma^2_{\text{noise}}$ \\
        \midrule
        1  & A & Normal         & 0.10 & 0.0  \\
        2  & A & \( t \)-distribution & 0.10 & 0.0  \\
        3  & A & \( t \)-distribution & 0.15 & 0.1  \\
        4  & B & \( t \)-distribution & 0.0  & 0.0  \\
        5  & B & Normal         & 0.0  & 0.0  \\
        6  & B & Normal         & 0.10 & 1.0  \\
        7 & C & \( t \)-distribution & 0.10 & 0.1 \\
        8 & C & Normal         & 0.10 & 0.1 \\
        9 & C & \( t \)-distribution & 0.0  & 1.0  \\
        \bottomrule
    \end{tabular}
\end{table}

\begin{table}[!ht]
\caption{Summary statistics of the number of FPC scores ($K$) that are selected in single, and bagging models during fitting for each scenarios.}
\label{tab:scenarios_K}
\begin{tabular*}{\textwidth}{@{\extracolsep\fill}lcccc}
\toprule
& \multicolumn{2}{@{}c@{}}{Single} & \multicolumn{2}{@{}c@{}}{Bagging} \\\cmidrule{2-3}\cmidrule{4-5}
Scenario & Range & Mean & Range & Mean \\
\hline 
1  & 3 -- 5  & 4.03  & 2 -- 6  & 3.88  \\
2  & 2 -- 5  & 3.91  & 1 -- 6  & 3.74 \\
3  & 2 -- 5  & 3.80  & 2 -- 6  & 3.68 \\
4  & 2 -- 6  & 4.15  & 1 -- 6  & 4.05 \\
5  & 3 -- 6  & 4.34  & 2 -- 6  & 4.21 \\
6  & 3 -- 5  & 3.74  & 2 -- 6  & 3.60 \\
7  & 2 -- 6  & 3.78  & 2 -- 6  & 3.62 \\
8  & 3 -- 5  & 3.85  & 2 -- 6  & 3.75 \\
9  & 2 -- 5  & 3.76   & 1 -- 6  & 3.72 \\
\hline 
\end{tabular*}
\end{table}

\subsection{Real data Analysis}
In this section, we apply the methods above to three real datasets, the Berkeley Growth Study dataset (\cite{tuddenham1954Physical}), spinal bone mineral density data (\cite{bachrach1999bone}), and the Electrocardiography (ECG) dataset (\cite{Robert2001}). 
\subsubsection{Berkeley Growth data}

\begin{figure}[H]
    \centering
    \includegraphics[width=0.85\linewidth]{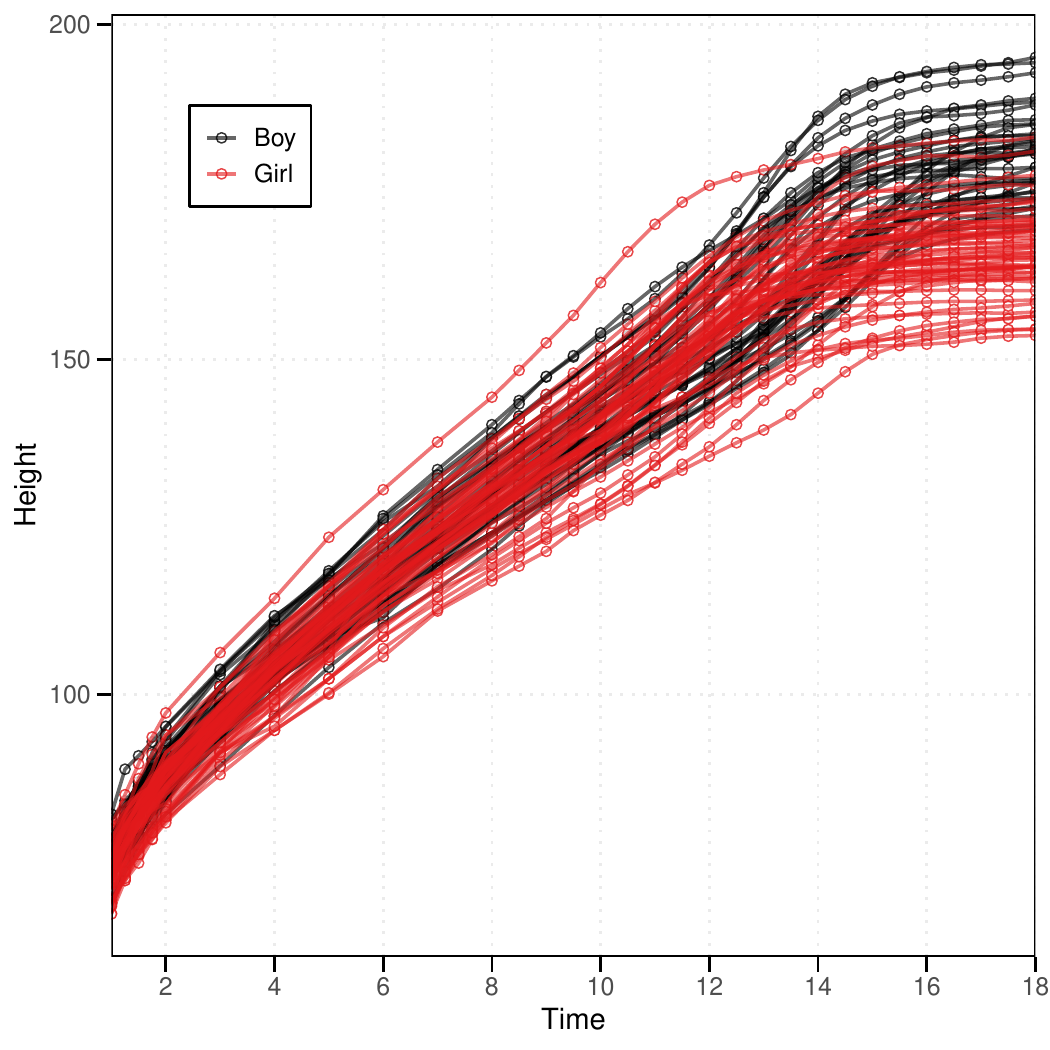}
    \caption{The Berkely growth data}
    \label{fig:growth_data}
\end{figure}

In this study, we analyze the Berkeley growth dataset (\cite{tuddenham1954Physical}), which contains height measurements for 93 individuals (54 girls and 39 boys). The dataset includes 31 time points, spanning ages 1 to 18, for each individual. The original growth curves are illustrated in Figure~\ref{fig:growth_data}.

To introduce sparsity artificially, we randomly reduce the number of observations per individual, selecting between 12 and 15 time points from the original set. For validation, we randomly split the 93 curves into a training set of 62 and a test set of 31, evaluating classification performance based on gender. The number of FPC scores, \( K \), is selected from a range of 2 to 5 for the bagging method and from a range of 3 to 4 for the single classifier approach. We select the number of FPC scores, $K$, by a PVE greater than
$0.99$. This process is repeated 500 times with different random splits, and the average classification results are summarized in Table~\ref{tab:berkeley}.

\subsubsection{Spinal bone mineral density data}
\begin{figure}[H]
    \centering
    \includegraphics[width=0.85\linewidth]{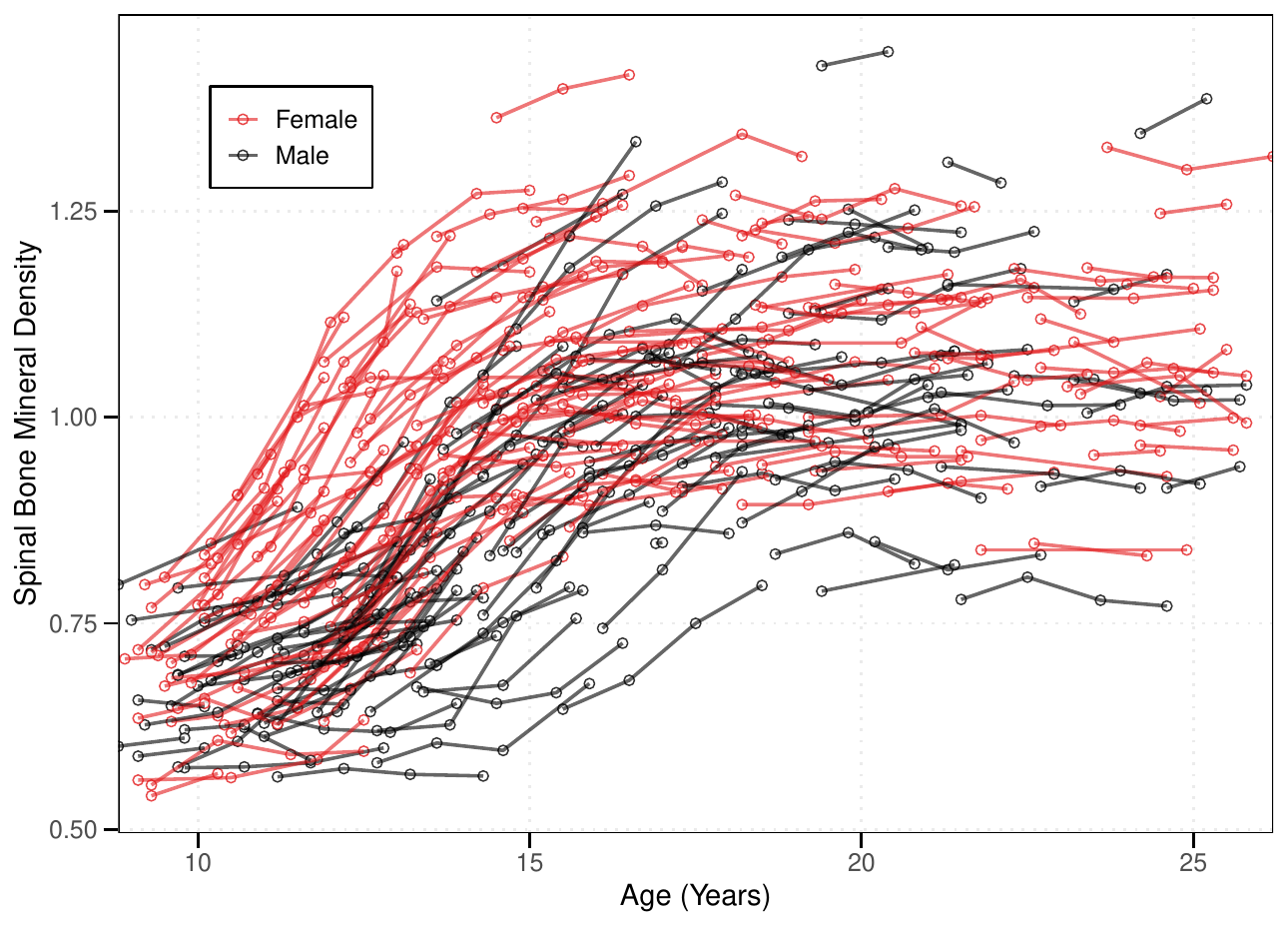}
    \caption{Spinal bone mineral density data}
    \label{fig:spnbmd_data}
\end{figure}

In this study, we analyze the spinal bone mineral density dataset (\cite{bachrach1999bone}), which contains measurements from 280 individuals (153 females and 127 males) recorded at sparse and irregular time points. Each individual has between two and four observations, as illustrated in Figure \ref{fig:spnbmd_data}. The primary objective is to evaluate various methods for gender classification.

To validate the models, we randomly split the dataset into 187 training and 93 test samples, applying the classification methods across 500 different data partitions. The number of FPC scores, \( K \), is selected from a range of 3 to 6 for the bagging method and from a range of 3 to 5 for the single classifier approach. We select the number of FPC scores, $K$, by a PVE greater than
$0.99$. The average classification results from these 500 splits are summarized in Table~\ref{tab:spinal}.

\subsubsection{Electrocardiography (ECG) data}

In this study, we analyze the ECG dataset (\cite{Robert2001}), which consists of 200 electrocardiogram (ECG) recordings, each representing the electrical activity of a single heartbeat. The dataset is categorized into two classes: 67 myocardial ECGs, characterized by supraventricular premature beats, and 133 normal ECGs. A plot of the ECG signals for each individual is shown in Figure \ref{fig:ECG}. Each signal was uniformly sampled at 96 time points to capture the temporal dynamics of the heartbeat. The primary objective is to evaluate various methods for the classification of ECGs.

Here, we artificially introduce sparsity into the data. We randomly select the number of observations for each individual from \{10, 11, 12\}, choosing the corresponding time points from the original set. For validation, we randomly split the dataset into 133 training and 67 test samples, assessing classification performance. The number of FPC scores, \( K \), is selected from a range of 3 to 6 for the bagging method and from a range of 4 to 5 for the single classifier approach. We select the number of FPC scores, $K$, by a PVE greater than
$0.99$. This process is repeated 500 times with different splits, and the average results are summarized in Table~\ref{tab:ECG}. 

\begin{figure}[H]
    \centering
    \includegraphics[width=0.85\linewidth]{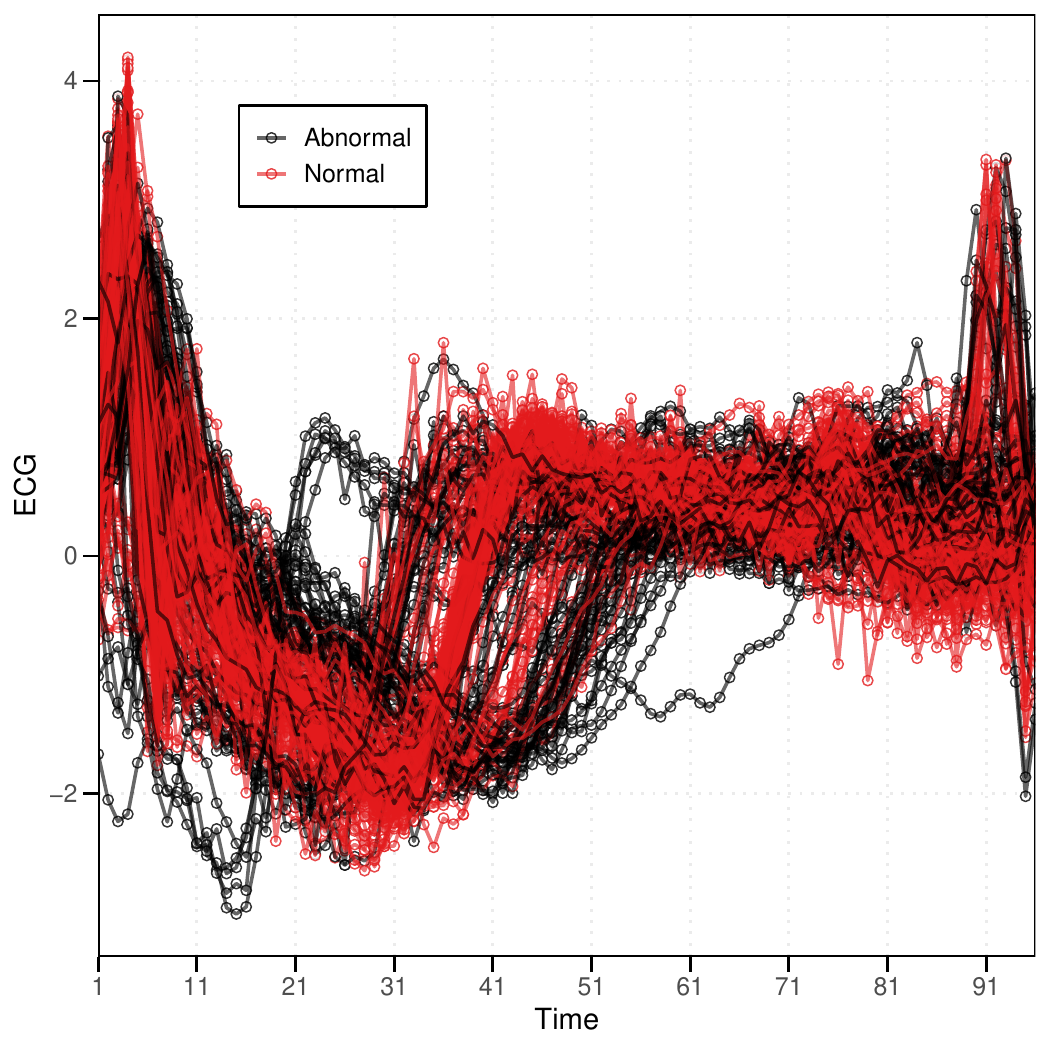}
    \caption{ECG data}
    \label{fig:ECG_data}
\end{figure}

\section{Results}\label{sec2}
\subsection{Simulation Results}

The performance of various classification methods across nine different scenarios is summarized in Tables~\ref{tab:modelA}, \ref{tab:modelB}, and \ref{tab:modelC}. Each table presents the average classification errors (\%) and their corresponding standard errors (in parentheses) computed from 500 Monte Carlo repetitions. The methods evaluated include logistic regression (Logit), linear discriminant analysis (LDA), quadratic discriminant analysis (QDA), naive Bayes, random forest (RF), and gradient boosting machines (GBM). Additionally, four aggregation strategies were employed: single, majority vote, out-of-bag (OOB) weight, and Bayesian.

In Scenarios 1--3, the Bayesian aggregation method consistently outperformed the other strategies across all classification methods. For example, in Scenario 1, the Bayesian approach achieved the lowest average classification error of {$19.16\%$} with RF, significantly lower than the single method's error of $24.40\%$. Similarly, in Scenarios 2 and 3, the Bayesian method with RF yielded the best results, with errors of {$18.27\%$} and {$19.78\%$}, respectively. This trend highlights the robustness of the Bayesian aggregation strategy in reducing classification errors, particularly when combined with ensemble methods like RF.

The majority vote and OOB weight methods also demonstrated improvements over the single method but were generally less effective than the Bayesian approach. For example, in Scenario 1, the majority vote method reduced the error of RF from $24.40\%$ to $20.22\%$, while the Bayesian method further reduced it to {$19.16\%$}. This suggests that while aggregation strategies generally improve performance, the Bayesian method provides the most significant enhancement.

Figures~\ref{fig:fig1}, \ref{fig:fig2}, \ref{fig:fig3}, \ref{fig:fig4}, \ref{fig:fig5}, \ref{fig:fig6}, \ref{fig:fig7}, \ref{fig:fig8}, and \ref{fig:fig9} in Appendix~\ref{appendix:simulation} show box plots visualizing the distribution of classification error rates for each classifier (GBM, LDA, Logit, Naïve Bayes, QDA, RF) over 500 simulation runs under four different methods: single, majority vote, out-of-bag (OOB) weighting, and Bayesian, for scenarios 1 to 9.

\begin{table}[h!]
    \centering
     \caption{The average classification errors (\%) and standard errors (in parentheses) from 500 monte carlo repetitions for the scenarios 1, 2, and 3}
    \label{tab:modelA}
    \begin{tabular}{llccccccc}
        \toprule
        Scenario & Method & Logit & LDA & QDA & NaiveBayes & RF & GBM \\
        \midrule
       1  & single         & 28.76 (7.91) & 28.52 (8.11) & 26.01 (7.93) & 25.87 (7.58) & 24.40 (6.90) & 25.64 (6.87) \\
            & Majority vote  & 25.96 (5.96) & 25.99 (6.11) & 23.59 (5.62) & 23.63 (5.41) & 20.22 (5.26) & 20.86 (5.48) \\
            & OOB weight     & 25.41 (5.86) & 25.40 (5.97) & 23.05 (5.51) & 23.16 (5.19) & 19.85 (5.27) & 20.42 (5.42) \\
            & Bayesian       & 24.51 (5.65) & 24.50 (5.65) & 22.10 (5.19) & 22.47 (4.96) & \bm{$19.16 (5.05)$} & 19.82 (5.28) \\
        \midrule
        2  & single         & 26.74 (7.90) & 26.66 (8.11) & 26.70 (8.57) & 27.00 (8.97) & 23.74 (6.78) & 24.52 (7.07) \\
            & Majority vote  & 23.28 (5.73) & 23.34 (5.89) & 22.90 (5.90) & 23.22 (5.95) & 19.10 (4.92) & 19.50 (4.97) \\
            & OOB weight     & 22.66 (5.62) & 22.63 (5.68) & 22.29 (5.65) & 22.37 (5.72) & 18.81 (4.85) & 19.18 (4.92) \\
            & Bayesian       & 21.87 (5.15) & 21.70 (5.17) & 21.69 (5.46) & 21.95 (5.56) & \bm{$18.27 (4.72)$} & 18.71 (4.98) \\
        \midrule
        3  & single         & 28.26 (8.34) & 28.52 (8.73) & 27.92 (8.53) & 27.99 (8.69) & 25.83 (7.06) & 26.43 (7.20) \\
            & Majority vote  & 25.01 (6.11) & 25.17 (6.39) & 24.73 (5.80) & 24.90 (5.96) & 20.73 (5.09) & 21.30 (5.15) \\
            & OOB weight     & 24.35 (5.90) & 24.46 (6.18) & 24.14 (5.63) & 24.06 (5.65) & 20.33 (4.94) & 20.95 (5.09) \\
            & Bayesian       & 23.22 (5.48) & 22.98 (5.45) & 23.43 (5.57) & 23.85 (5.44) & \bm{$19.78 (4.77)$} & 20.29 (5.00) \\
        \bottomrule
    \end{tabular}
\end{table}

Scenarios 4--6 exhibit a similar trend, with the Bayesian method consistently delivering the lowest classification errors. In Scenario 4, the Bayesian approach with LDA achieved the lowest error of {$8.62\%$}, compared to $10.05\%$ for the single method. Similarly, in Scenario 5, the Bayesian method with LDA resulted in an error of {$9.81\%$}, outperforming the single method's error of $10.76\%$. In Scenario 6, the Bayesian method with Logit achieved the lowest error of {$26.07\%$}, demonstrating its versatility across different classification techniques.

The majority vote and OOB weight methods again provided intermediate performance improvements. For example, in Scenario 4, the majority vote method reduced the error of LDA from $10.05\%$ to $9.00\%$, while the Bayesian method further reduced it to {$8.62\%$}. This reinforces the superiority of the Bayesian approach in minimizing classification errors.

\begin{table}[h!]
    \centering
        \caption{The average classification errors (\%) and standard errors (in parentheses) from 500 monte carlo repetitions for the scenarios 4, 5, and 6}
    \label{tab:modelB}
    \begin{tabular}{llccccccc}
        \toprule
        Scenario & Method & Logit & LDA & QDA & NaiveBayes & RF & GBM \\
        \midrule
        4  & single         & 10.59 (4.24) & 10.05 (4.21) & 11.71 (5.12) & 12.56 (5.45) & 12.87 (4.16) & 12.80 (4.11) \\
            & Majority vote  & 9.26 (3.03) & 9.00 (2.92) & 10.08 (3.20) & 10.72 (3.39) & 10.51 (3.39) & 10.25 (3.34) \\
            & OOB weight     & 9.17 (3.05) & 8.79 (2.86) & 9.91 (3.17) & 10.38 (3.21) & 10.38 (3.39) & 10.12 (3.36) \\
            & Bayesian       & 8.77 (2.99) & \bm{$8.62 (2.90)$} & 9.51 (2.99) & 10.08 (3.35) & 10.13 (3.36) & 9.90 (3.34) \\
        \midrule
        5  & single         & 11.31 (3.49) & 10.76 (3.47) & 12.26 (3.85) & 13.58 (4.16) & 14.60 (3.96) & 14.77 (3.97) \\
            & Majority vote  & 10.37 (3.29) & 10.18 (3.28) & 11.21 (3.46) & 12.25 (3.73) & 12.31 (3.60) & 12.12 (3.59) \\
            & OOB weight     & 10.32 (3.30) & 10.09 (3.24) & 11.07 (3.44) & 11.90 (3.55) & 12.16 (3.59) & 12.05 (3.52) \\
            & Bayesian       & 10.02 (3.24) & \bm{$9.81 (3.13)$} & 10.61 (3.21) & 11.58 (3.48) & 11.64 (3.44) & 11.60 (3.49) \\
        \midrule
        6  & single         & 28.63 (5.54) & 28.56 (5.58) & 31.06 (5.75) & 30.56 (5.84) & 30.81 (5.53) & 30.60 (5.81) \\
            & Majority vote  & 27.21 (5.20) & 27.20 (5.19) & 29.47 (5.41) & 28.98 (5.31) & 27.31 (4.93) & 27.00 (5.01) \\
            & OOB weight     & 27.02 (5.08) & 27.00 (5.19) & 29.16 (5.28) & 28.69 (5.29) & 27.06 (4.93) & 26.85 (4.98) \\
            & Bayesian       & \bm{$26.07 (4.90)$} & 26.05 (4.95) & 28.21 (5.28) & 27.80 (5.11) & 26.33 (4.65) & 26.23 (4.80) \\
        \bottomrule
    \end{tabular}
\end{table}

In Scenarios 7--9, the Bayesian method continued to excel, particularly with Naive Bayes, achieving the lowest errors of {$36.12\%$}, {$31.61\%$}, and {$38.05\%$} in Scenarios 7, 8, and 9, respectively. These results are notably lower than those obtained with the single method, which had errors of $38.41\%$, $34.14\%$, and $40.25\%$ in the respective scenarios. The Bayesian method's ability to leverage probabilistic weighting likely contributed to its superior performance.

The majority vote and OOB weight methods again provided moderate improvements. For example, in Scenario 7, the majority vote method reduced the error of Naive Bayes from $38.41\%$ to $37.35\%$, while the Bayesian method further reduced it to {$36.12\%$}. This pattern underscores the Bayesian method's effectiveness in refining classification outcomes.

\begin{table}[h!]
    \centering
        \caption{The average classification errors (\%) and standard errors (in parentheses) from 500 monte carlo repetitions for the scenarios 7, 8, and 9.}
    \label{tab:modelC}
    \begin{tabular}{llccccccc}
        \toprule
        Scenario & Method & Logit & LDA & QDA & NaiveBayes & RF & GBM \\
        \midrule
        7 & single         & 50.46 (5.68) & 50.51 (5.64) & 38.62 (5.83) & 38.41 (5.76) & 41.98 (5.42) & 43.71 (5.43) \\
            & Majority vote  & 50.38 (5.81) & 50.40 (5.86) & 37.56 (5.40) & 37.35 (5.48) & 39.68 (5.17) & 40.73 (5.29) \\
            & OOB weight     & 50.14 (5.77) & 50.26 (5.81) & 37.44 (5.36) & 37.23 (5.38) & 39.62 (5.21) & 40.71 (5.20) \\
            & Bayesian       & 45.34 (4.43) & 45.35 (4.44) & 36.25 (5.37) & \bm{$36.12 (5.11)$} & 38.01 (4.95) & 39.12 (4.83) \\
        \midrule
        8 & single         & 50.53 (5.97) & 50.56 (5.95) & 34.34 (5.57) & 34.14 (5.64) & 37.38 (5.74) & 40.41 (5.70) \\
            & Majority vote  & 50.08 (6.13) & 50.14 (6.25) & 33.45 (5.05) & 33.10 (5.30) & 34.78 (5.33) & 36.82 (5.64) \\
            & OOB weight     & 49.77 (6.08) & 49.77 (6.14) & 33.35 (4.93) & 32.98 (5.11) & 34.76 (5.30) & 36.83 (5.65) \\
            & Bayesian       & 45.13 (4.76) & 45.18 (4.92) & 31.97 (5.05) & \bm{$31.61 (4.87)$} & 33.26 (5.00) & 34.96 (5.27) \\
        \midrule
       9 & single         & 51.17 (5.33) & 51.14 (5.34) & 41.14 (5.51) & 40.25 (5.43) & 44.40 (5.07) & 45.69 (5.25) \\
            & Majority vote  & 50.81 (5.15) & 50.78 (5.25) & 40.54 (5.65) & 39.69 (5.67) & 42.68 (5.39) & 43.58 (5.61) \\
            & OOB weight     & 50.73 (5.20) & 50.70 (5.21) & 40.55 (5.61) & 39.53 (5.57) & 42.66 (5.43) & 43.55 (5.46) \\
            & Bayesian       & 45.22 (3.92) & 45.24 (3.93) & 39.03 (4.84) & \bm{$38.05 (5.00)$} & 40.65 (4.95) & 41.55 (4.75) \\
        \bottomrule
    \end{tabular}
\end{table}

Across all scenarios, the Bayesian aggregation method consistently delivered the lowest classification errors, demonstrating its robustness and effectiveness. The majority vote and OOB weight methods also improved performance compared to the single method but were generally less effective than the Bayesian approach. 

\begin{figure}[H]
    \centering
    \includegraphics[width=0.85\linewidth]{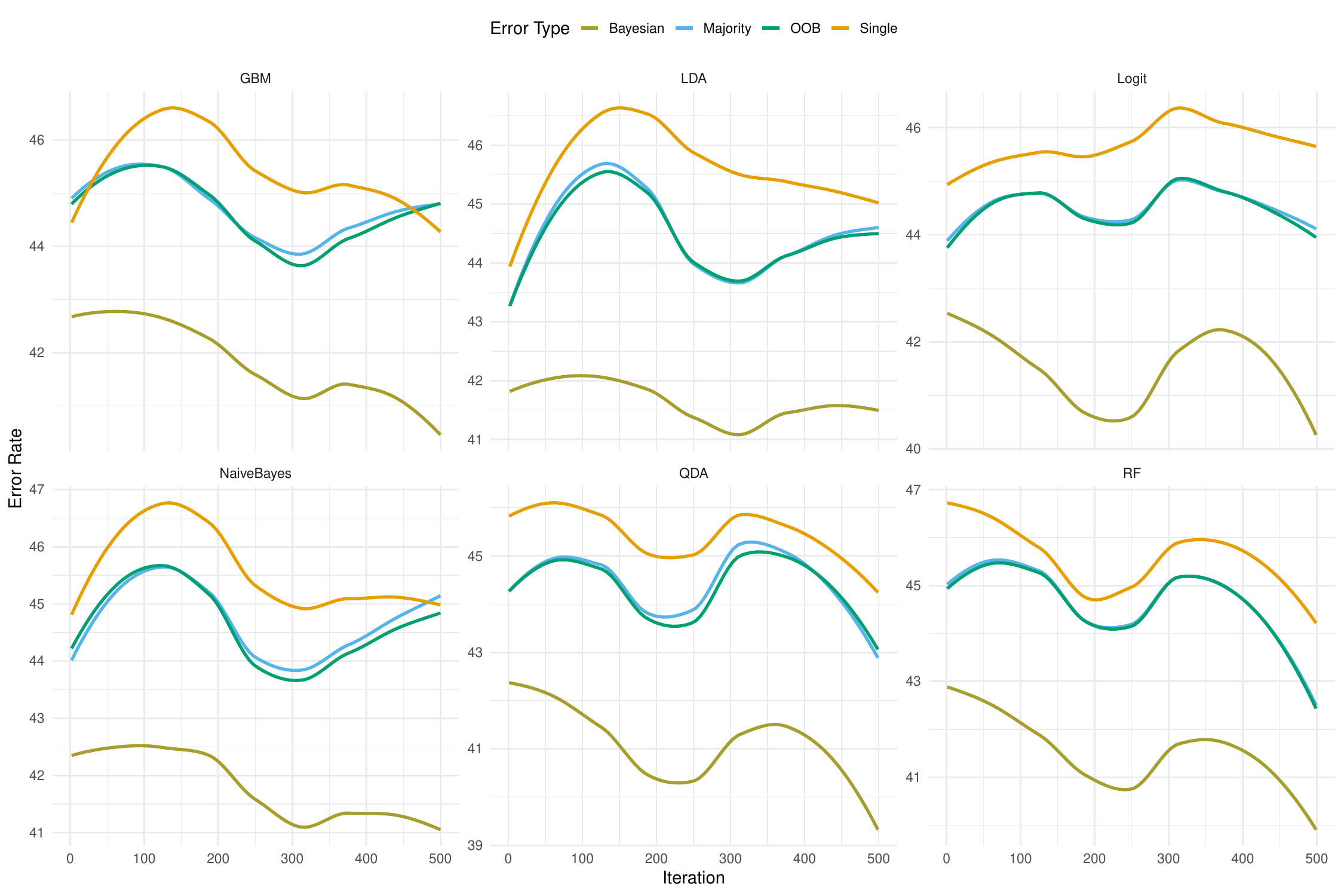}
\caption{Error rate trends over 500 iterations for different classifiers (Logit, LDA, QDA, Naïve Bayes, RF, GBM) under four classification methods (single, majority vote, OOB weight, Bayesian) in Scenario 9.}    
\label{fig:error_rate_7}
\end{figure}

Figure~\ref{fig:error_rate_7} illustrates how classification errors in Scenario 9 fluctuate in iterations, highlighting the stability and performance of the Bayesian approach.

\subsection{Real Data Results}

The findings from 500 random splits in the Berkeley growth dataset in Table~\ref{tab:berkeley} underscore the efficacy of the proposed Bayesian model in achieving superior classification accuracy. Notably, the Bayesian approach attained the lowest classification error of {4.03\%} with QDA, outperforming all alternative methods. Its consistent advantage over single classifiers and other ensemble techniques highlights its robustness and effectiveness in minimizing misclassification. The results affirm that Bayesian aggregation not only refines predictive precision but also enhances model stability across varying data partitions, reinforcing its suitability for complex classification tasks.

\begin{table}[h!]
    \centering
    \caption{The average classification errors (\%) and standard errors (in parentheses) from 500 random
    splits in the Berkeley growth data.}
    \label{tab:berkeley}
    
    \begin{tabular}{lccccccc}
        \toprule
        Method & Logit & LDA & QDA & NaiveBayes & RF & GBM \\
        \midrule
        single         & 7.47 (4.48)  & 6.01 (3.61)  & 6.93 (4.04)  & 7.47 (4.66)  & 7.95 (4.73)  & 9.93 (5.35)  \\
        Majority vote  & 5.58 (3.49)  & 5.14 (3.23)  & 4.87 (3.12)  & 5.80 (3.79)  & 6.24 (3.82)  & 8.31 (4.56)  \\
        OOB weight     & 5.64 (3.47)  & 5.21 (3.17)  & 4.91 (3.14)  & 5.55 (3.57)  & 6.13 (3.76)  & 7.85 (4.42)  \\
        Bayesian       & 4.67 (3.46)  & 4.21 (3.24)  & \textbf{4.03 (3.22)}  & 5.01 (3.78)  & 5.42 (3.71)  & 6.98 (4.49)  \\
        \bottomrule
    \end{tabular}

\end{table}

\begin{figure}[H]
    \centering
    \includegraphics[width=1\linewidth]{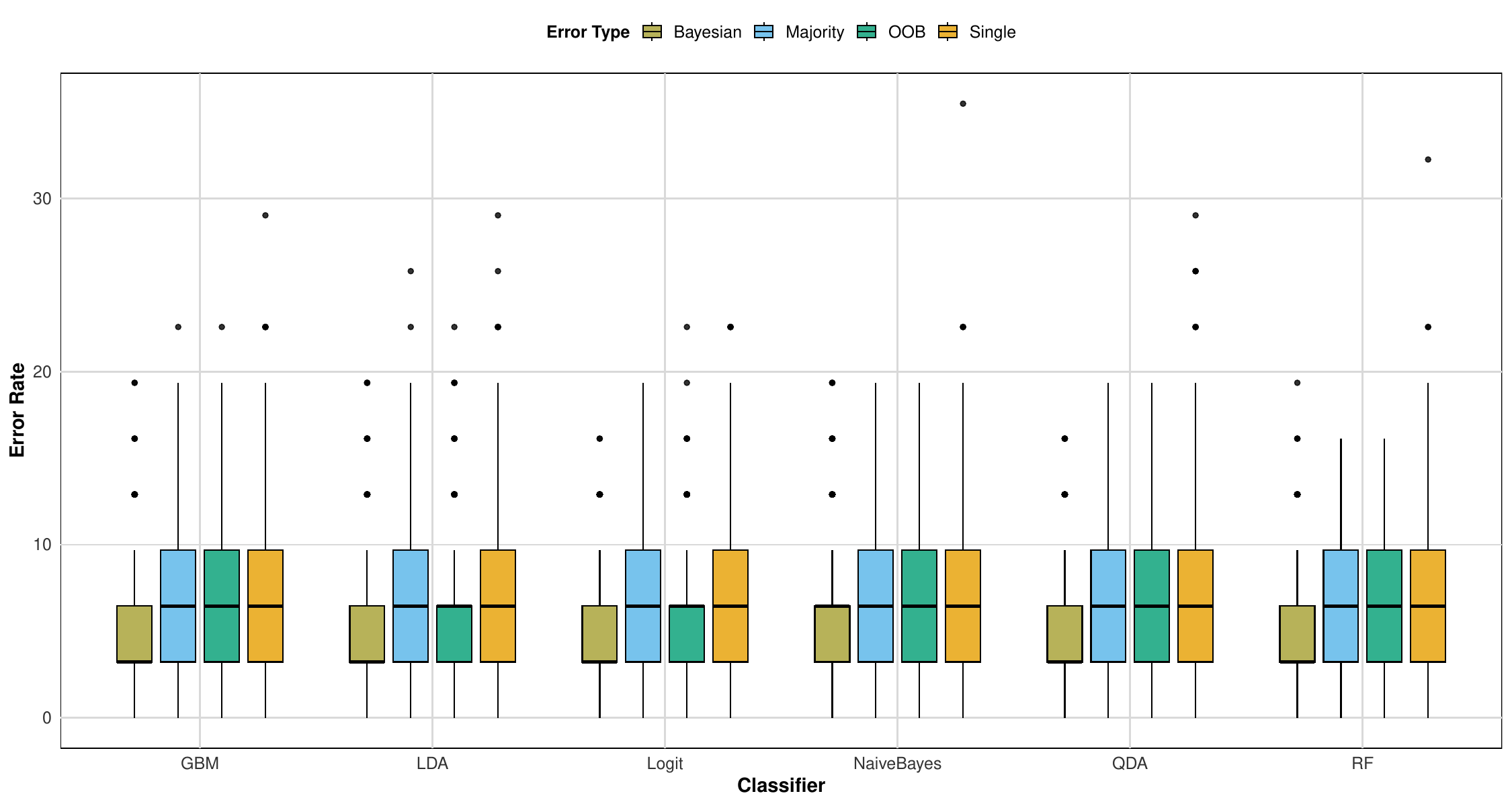}
 \caption{Boxplot of classification error rates across different classifiers (GBM, LDA, Logit, Naïve Bayes, QDA, RF) under four different methods: single, majority vote, OOB weighting, and Bayesian in Berkeley growth data. Each boxplot represents the distribution of classification errors over 500 simulations}
    \label{fig:berkeley}
\end{figure}

Figure~\ref{fig:berkeley} shows a box plot that visualizes the distribution of classification error rates for each classifier (GBM, LDA, Logit, Nave Bayes, QDA, RF) in 500 simulation runs under four different methods: single, majority vote, out-of-bag (OOB) weighting, and Bayesian in the Berkeley growth data. Notably, the Bayesian method demonstrates both lower median error rates and reduced variability compared to alternative approaches, reinforcing its robustness and stability. In contrast, the single classifier method exhibits higher median errors and greater dispersion, indicating less reliable performance. These results confirm that Bayesian aggregation improves classification accuracy and enhances model consistency across multiple data partitions, making it a compelling choice for robust predictive modeling.

The results from 500 random splits of the spinal bone mineral density dataset in Table~\ref{tab:spinal} highlight the effectiveness of the proposed Bayesian model in classification accuracy. The Bayesian approach with LDA achieved the lowest classification error of {29.95\% }, demonstrating its superiority over alternative methods. Compared to single classifiers and other ensemble techniques, the Bayesian model consistently exhibited enhanced performance, reducing misclassification rates while maintaining robustness across different data partitions. These findings reinforce the efficacy of Bayesian aggregation in improving predictive accuracy and model stability in complex classification scenarios.

\begin{table}[h!]
    \centering
    \caption{The average classification errors (\%) and standard errors (in parentheses) from 500 random
    splits of the spinal bone mineral density data.}
    \label{tab:spinal}
    \begin{tabular}{lccccccc}
        \toprule
        Method & Logit & LDA & QDA & NaiveBayes & RF & GBM \\
        \midrule
        single         & 32.02 (4.21)  & 31.96 (4.22)  & 35.73 (4.56)  & 34.85 (5.07)  & 36.00 (4.57)  & 34.84 (4.96)  \\
        Majority vote  & 31.12 (3.99)  & 31.11 (3.98)  & 32.39 (4.22)  & 32.28 (4.28)  & 32.41 (4.13)  & 32.55 (4.13)  \\
        OOB weight     & 31.14 (3.96)  & 31.19 (3.91)  & 32.37 (4.21)  & 32.30 (4.23)  & 32.43 (4.16)  & 32.55 (4.11)  \\
        Bayesian       & 29.95 (4.08)  & \bm{$29.95 (3.97)$}  & 31.54 (4.41)  & 31.42 (4.41)  & 31.52 (4.17)  & 31.46 (3.98)  \\
        \bottomrule
    \end{tabular}
\end{table}

\begin{figure}[H]
    \centering
    \includegraphics[width=1\linewidth]{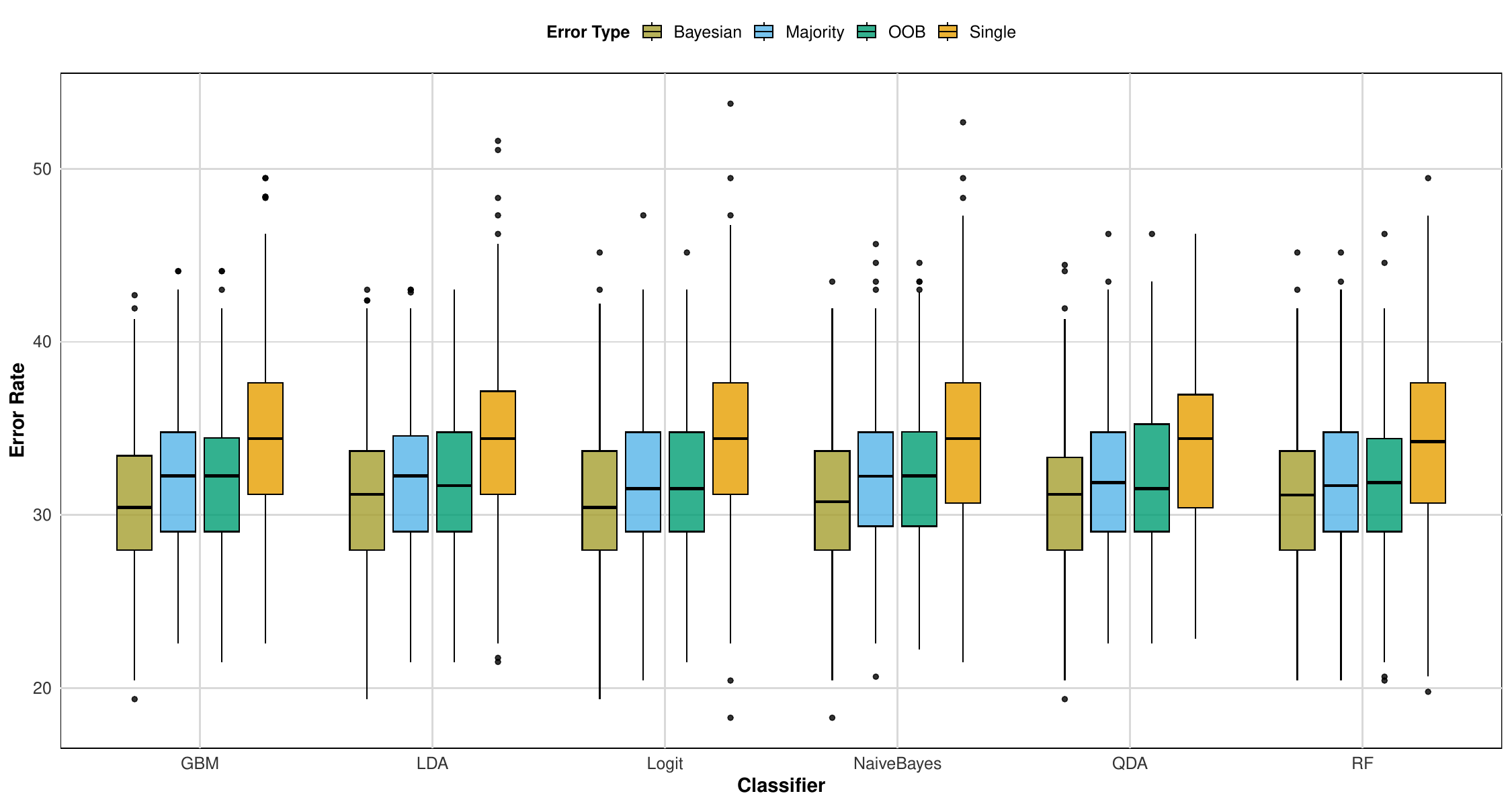}
 \caption{Boxplot of classification error rates across different classifiers (GBM, LDA, Logit, Naïve Bayes, QDA, RF) under four different methods: single, majority vote, OOB weighting, and Bayesian in the spinal bone mineral density data. Each boxplot represents the distribution of classification errors over 500 simulations}
    \label{fig:spnbmd}
\end{figure}

Figure~\ref{fig:spnbmd} show a boxplot that presents the distribution of classification error rates across different classifiers (GBM, LDA, Logit, Naïve Bayes, QDA, RF) under four distinct methods: single, majority vote, out-of-bag (OOB) weighting, and Bayesian in the spinal bone mineral density data. The Bayesian method demonstrates a noticeable reduction in both median classification error and variability, particularly for GBM and Logit, reinforcing its advantage in achieving more stable and accurate classification. The single classifier method consistently exhibits higher median error rates and wider variability, indicating less reliable performance across different data partitions. Meanwhile, ensemble techniques such as majority vote and OOB weighting reduce classification errors compared to the single method but show greater variability than Bayesian aggregation.

The results from 500 random splits of the ECG dataset in Table~\ref{tab:ECG} underscore the superior performance of the proposed Bayesian model in classification accuracy. Notably, the Bayesian approach with RF achieved the lowest classification error of {20.42\%}, outperforming all other methods. Compared to single classifiers and other ensemble techniques, the Bayesian model demonstrated a consistent reduction in misclassification rates, reinforcing its robustness across varying data partitions. These findings highlight the effectiveness of Bayesian aggregation in refining predictive precision and improving model stability, making it a compelling choice for ECG signal classification.

\begin{table}[h!]
    \centering
    \caption{The average classification errors (\%) and standard errors (in parentheses) from 500 random
splits of ECG data.}
    \label{tab:ECG}
    \begin{tabular}{lccccccc}
        \toprule
        Method & Logit & LDA & QDA & NaiveBayes & RF & GBM \\
        \midrule
        single         & 22.11 (4.17)  & 22.16 (4.07)  & 21.77 (4.52)  & 22.91 (4.42)  & 22.45 (4.14)  & 22.92 (4.49)  \\
        Majority vote  & 22.10 (4.34)  & 22.08 (4.30)  & 22.26 (4.48)  & 22.44 (4.30)  & 21.28 (4.29)  & 21.75 (4.48)  \\
        OOB weight     & 22.13 (4.30)  & 22.07 (4.31)  & 22.15 (4.43)  & 22.41 (4.30)  & 21.32 (4.32)  & 21.74 (4.52)  \\
        Bayesian       & 21.39 (4.19)  & 21.46 (4.25)  & 21.44 (4.46)  & 21.78 (4.24)  & \bm{$20.42 (4.13)$}  & 20.97 (4.26)  \\
        \bottomrule
    \end{tabular}
\end{table}

\begin{figure}[H]
    \centering
    \includegraphics[width=1\linewidth]{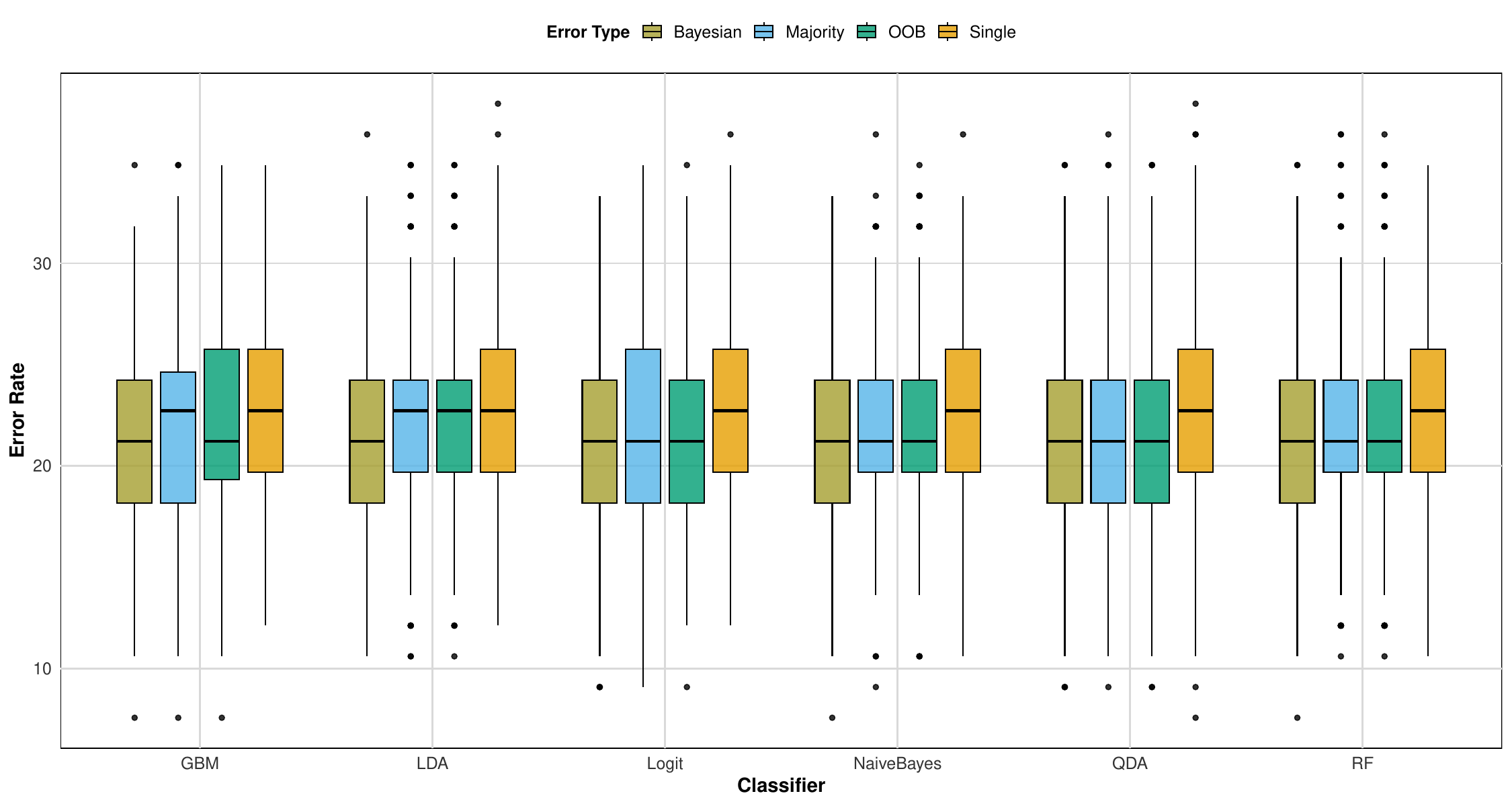}
 \caption{Boxplot of classification error rates across different classifiers (GBM, LDA, Logit, Naïve Bayes, QDA, RF) under four different methods: single, majority vote, OOB weighting, and Bayesian in the ECG dataset. Each boxplot represents the distribution of classification errors over 500 simulations}
    \label{fig:ECG}
\end{figure}

Figure~\ref{fig:ECG} shows a boxplot visualizes the classification error distributions across different classifiers (GBM, LDA, Logit, Naïve Bayes, QDA, RF) evaluated under four method frameworks: single, majority vote, out-of-bag (OOB) weighting, and Bayesian in the ECG dataset. Among the methods, Bayesian aggregation consistently exhibits lower median classification errors and reduced variability, highlighting its robustness in ECG classification. In particular, the Bayesian model with all classifiers achieve the lowest error rates with minimal dispersion, strengthening its effectiveness in capturing the underlying patterns of ECG signals. In contrast, the single classifier method consistently shows higher error rates and wider variability, indicating its sensitivity to different data partitions. The majority vote and OOB weighting methods provide moderate improvements over single classifiers but do not match the stability and accuracy of the Bayesian approach.

\section{Conclusion}

In this study, we introduced a novel classification method for sparse functional data that integrates functional principal component analysis (FPCA) with Bayesian aggregation. Unlike traditional ensemble techniques that rely on majority voting or out-of-bag error (OOB) weight, our Bayesian aggregation framework refines classifier predictions by leveraging Bayesian calibration, explicitly modeling uncertainty, and producing well-calibrated probability estimates. Through extensive simulation studies and real-world applications, we demonstrated the superior performance of the proposed method compared to conventional single classifiers and other ensemble techniques.

Our results highlight the effectiveness of Bayesian aggregation in enhancing classification accuracy across diverse scenarios, particularly in the presence of data sparsity and irregular sampling. In simulation studies, the Bayesian aggregation method consistently achieved the lowest classification errors across multiple settings, outperforming both majority voting and OOB weighting techniques. Furthermore, the method exhibited remarkable robustness in handling outlier contamination and noise, underscoring its practical utility in real-world applications.

The real-data analyses further validated the efficacy of our approach. Across three distinct datasets--the Berkeley Growth Study, the spinal bone mineral density dataset, and the electrocardiography (ECG) dataset--Bayesian aggregation consistently delivered lower classification errors and greater stability compared to alternative methods. These results underscore the advantages of incorporating Bayesian inference into ensemble learning for functional data classification.

Beyond improving classification accuracy, this study makes broader method contributions. By integrating Bayesian principles with functional data analysis, we provide a principled framework for addressing uncertainty in classifier outputs. This approach offers valuable insights for domains where reliable classification is critical, such as healthcare, genomics, and time-series forecasting.

Future research directions include extending the Bayesian aggregation framework to multi-class classification problems, exploring alternative prior structures for Bayesian calibration.

In conclusion, the proposed Bayesian aggregation method represents an advancement in the classification of sparse functional data. By effectively combining ensemble learning with Bayesian inference, our approach delivers improved accuracy, robustness, and interpretability, making it a tool for analyzing complex functional datasets.

\begin{appendices}

\section{Simulation Results}
\label{appendix:simulation}

\begin{figure}[H]
    \centering
\includegraphics[width=0.80\textwidth]{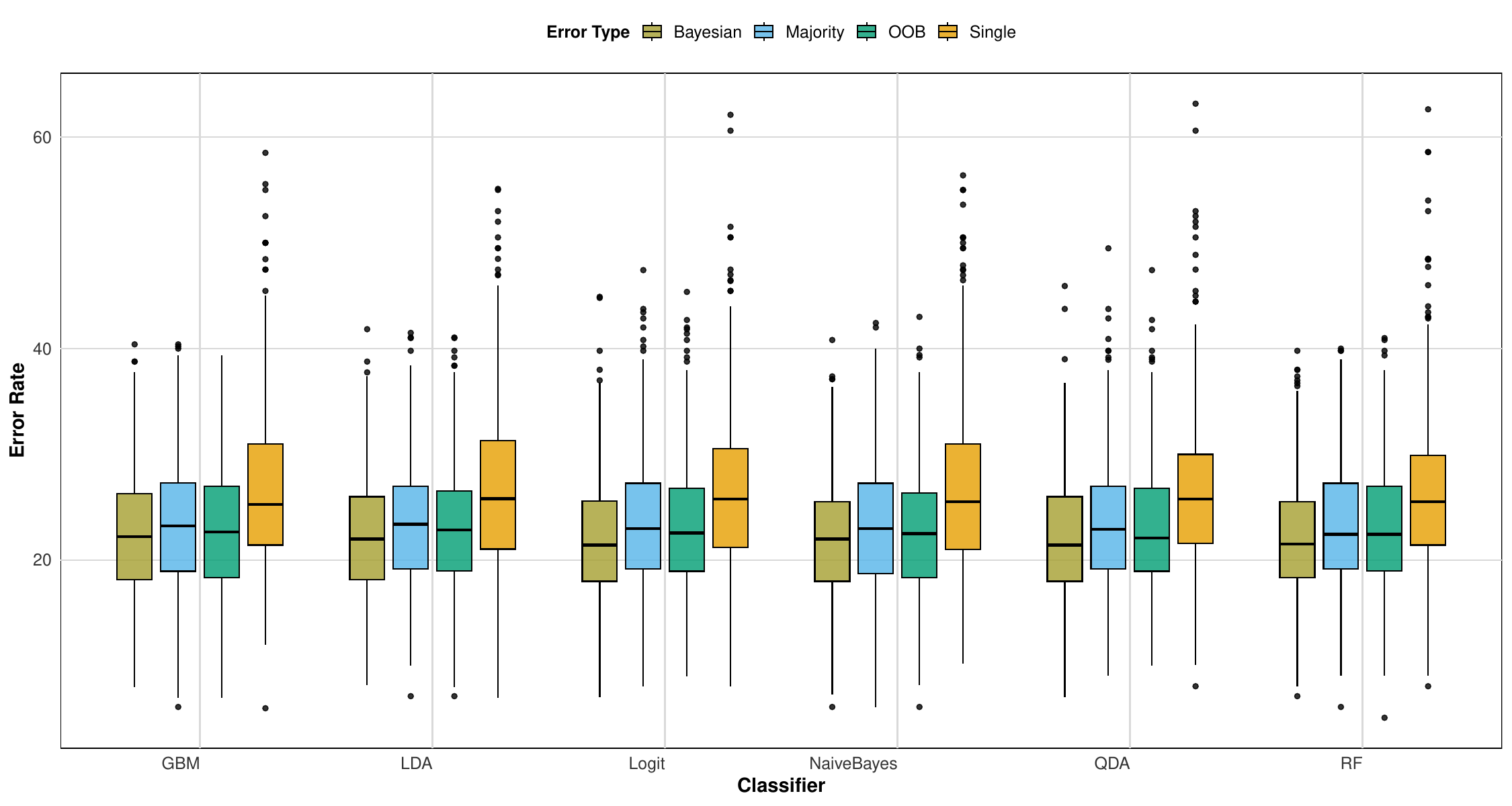}
    \caption{Boxplot of error rates for different classifiers (Logit, LDA, QDA, Naïve Bayes, RF, GBM) across four approaches (single, majority vote, OOB weight, Bayesian) for scenario 1 setting. Each boxplot represents the distribution of error rates computed from 500 replicated simulations.}
    \label{fig:fig1}
\end{figure}

\begin{figure}[H]
    \centering
    \includegraphics[width=0.80\textwidth]{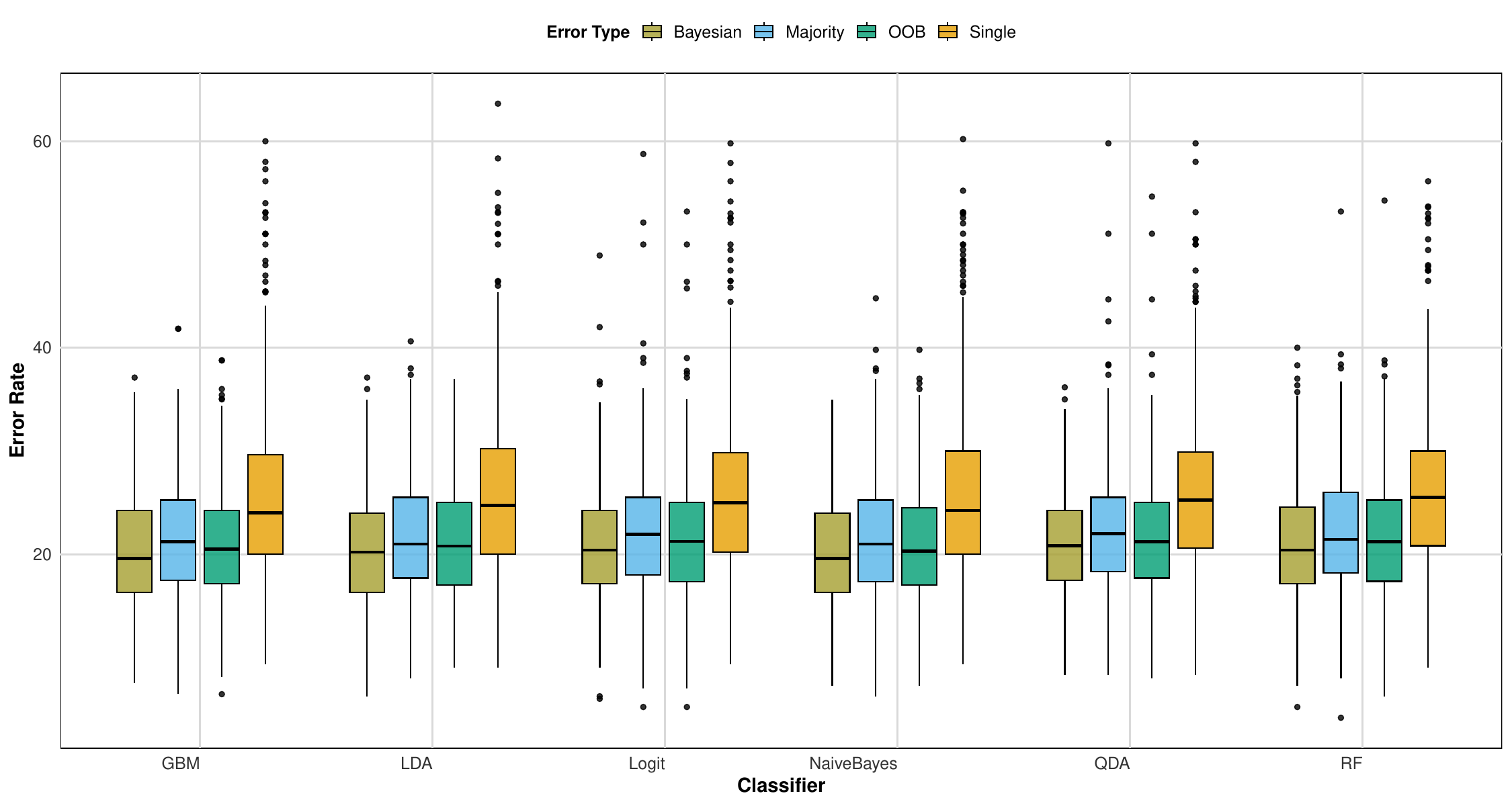}
    \caption{Boxplot of error rates for different classifiers (Logit, LDA, QDA, Naïve Bayes, RF, GBM) across four approaches (single, majority vote, OOB weight, Bayesian) for scenario 2 setting. Each boxplot represents the distribution of error rates computed from 500 replicated simulations.}
    \label{fig:fig2}
\end{figure}

\begin{figure}[H]
    \centering
    \includegraphics[width=0.80\textwidth]{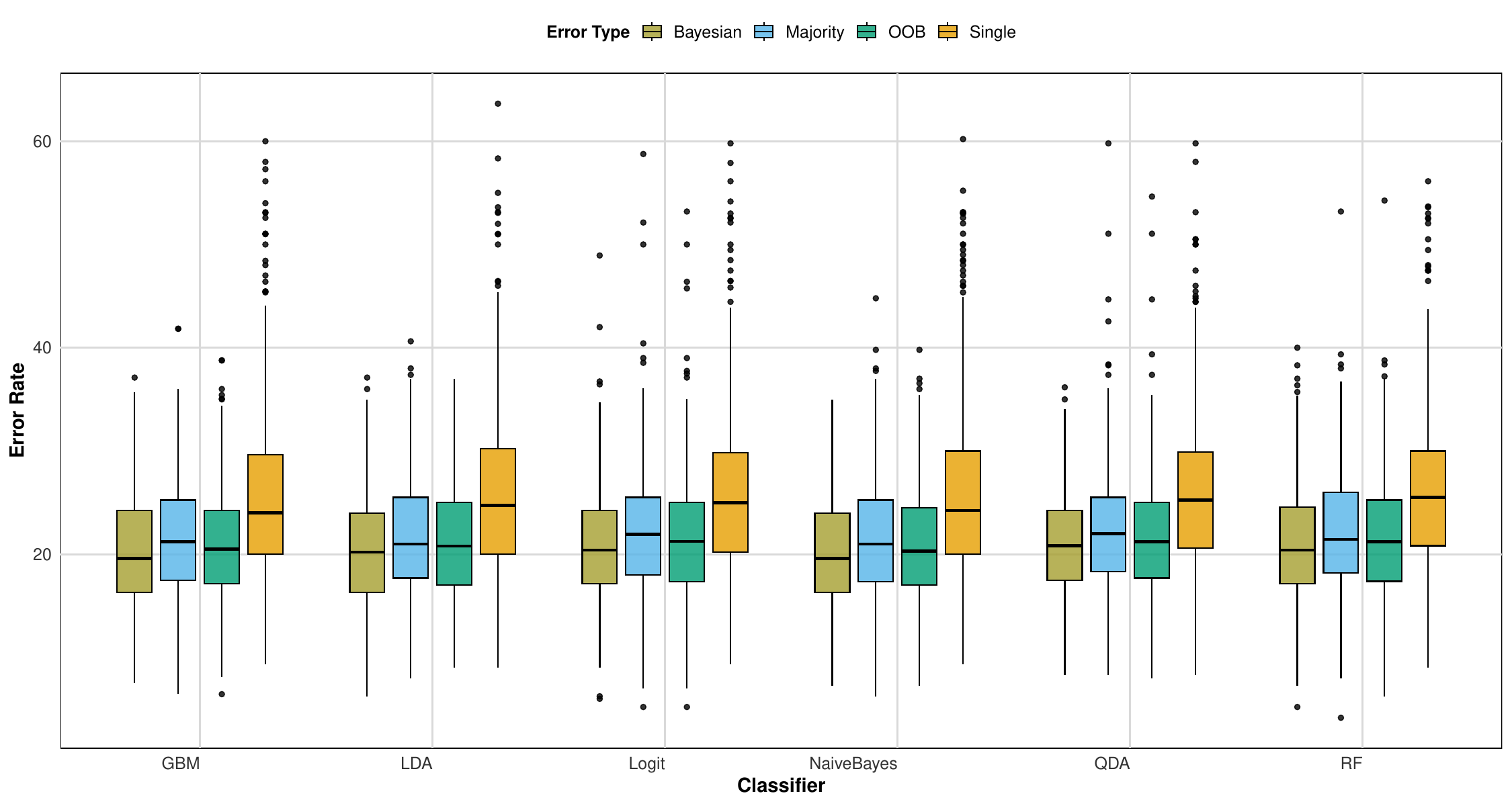}
    \caption{Boxplot of error rates for different classifiers (Logit, LDA, QDA, Naïve Bayes, RF, GBM) across four approaches (single, majority vote, OOB weight, Bayesian) for scenario 3 setting. Each boxplot represents the distribution of error rates computed from 500 replicated simulations.}
    \label{fig:fig3}
\end{figure}

\begin{figure}[H]
    \centering
    \includegraphics[width=0.80\textwidth]{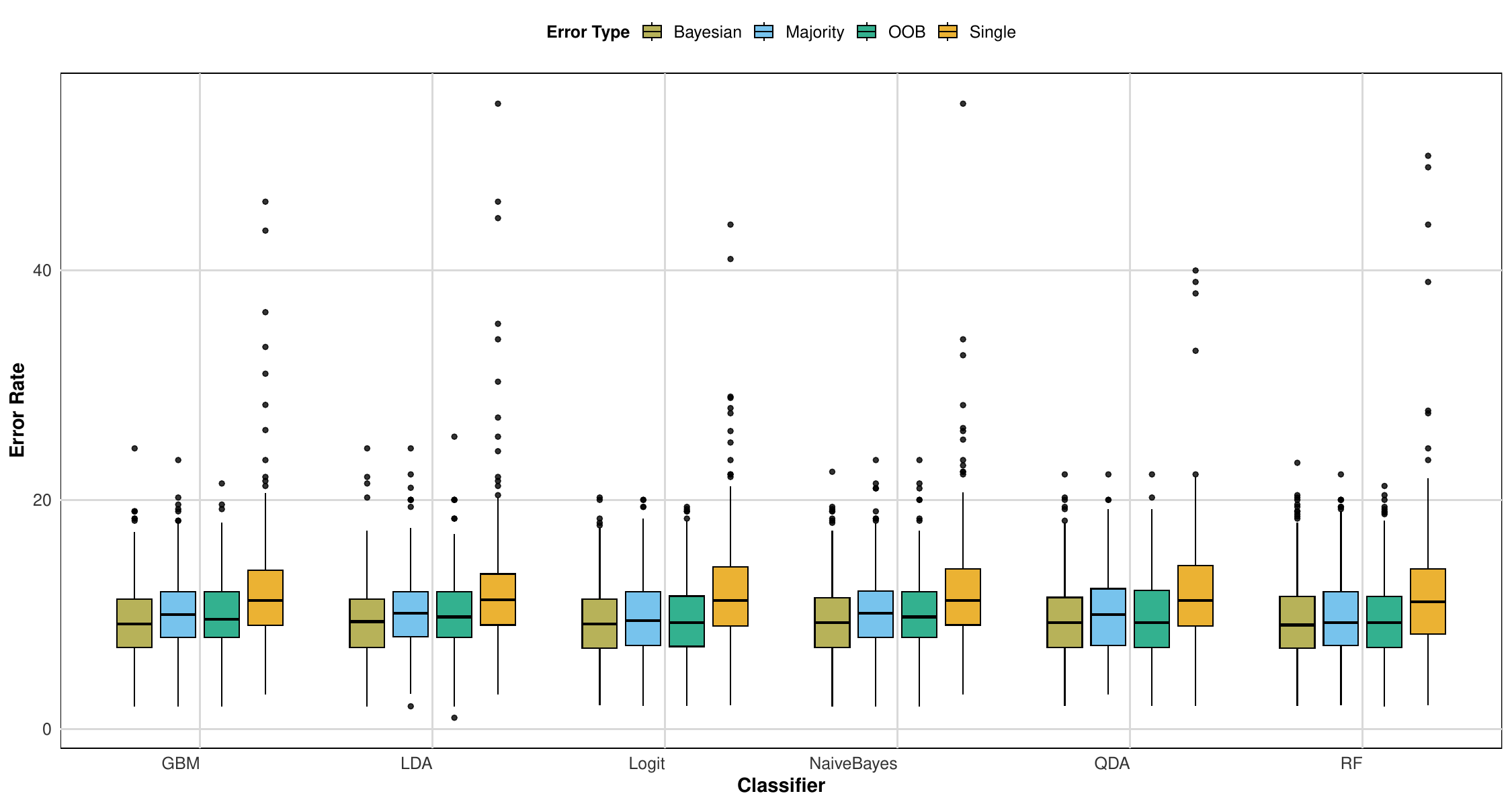}
    \caption{Boxplot of error rates for different classifiers (Logit, LDA, QDA, Naïve Bayes, RF, GBM) across four approaches (single, majority vote, OOB weight, Bayesian) for scenario 4 setting. Each boxplot represents the distribution of error rates computed from 500 replicated simulations.}
    \label{fig:fig4}
\end{figure}

\begin{figure}[H]
    \centering
    \includegraphics[width=0.80\textwidth]{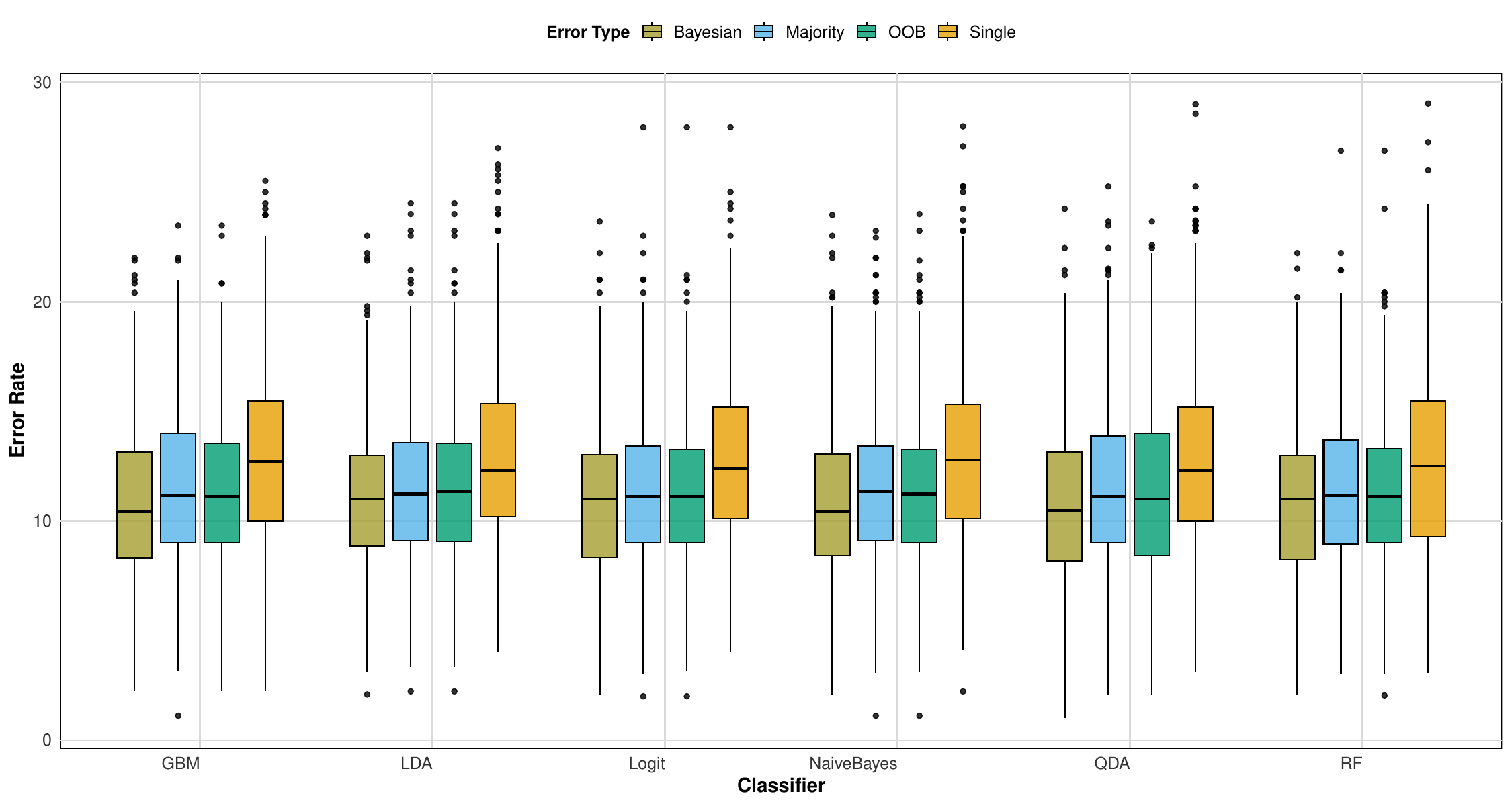}
    \caption{Boxplot of error rates for different classifiers (Logit, LDA, QDA, Naïve Bayes, RF, GBM) across four approaches (single, majority vote, OOB weight, Bayesian) for scenario 5 setting. Each boxplot represents the distribution of error rates computed from 500 replicated simulations.}
    \label{fig:fig5}
\end{figure}

\begin{figure}[H]
    \centering
    \includegraphics[width=0.80\textwidth]{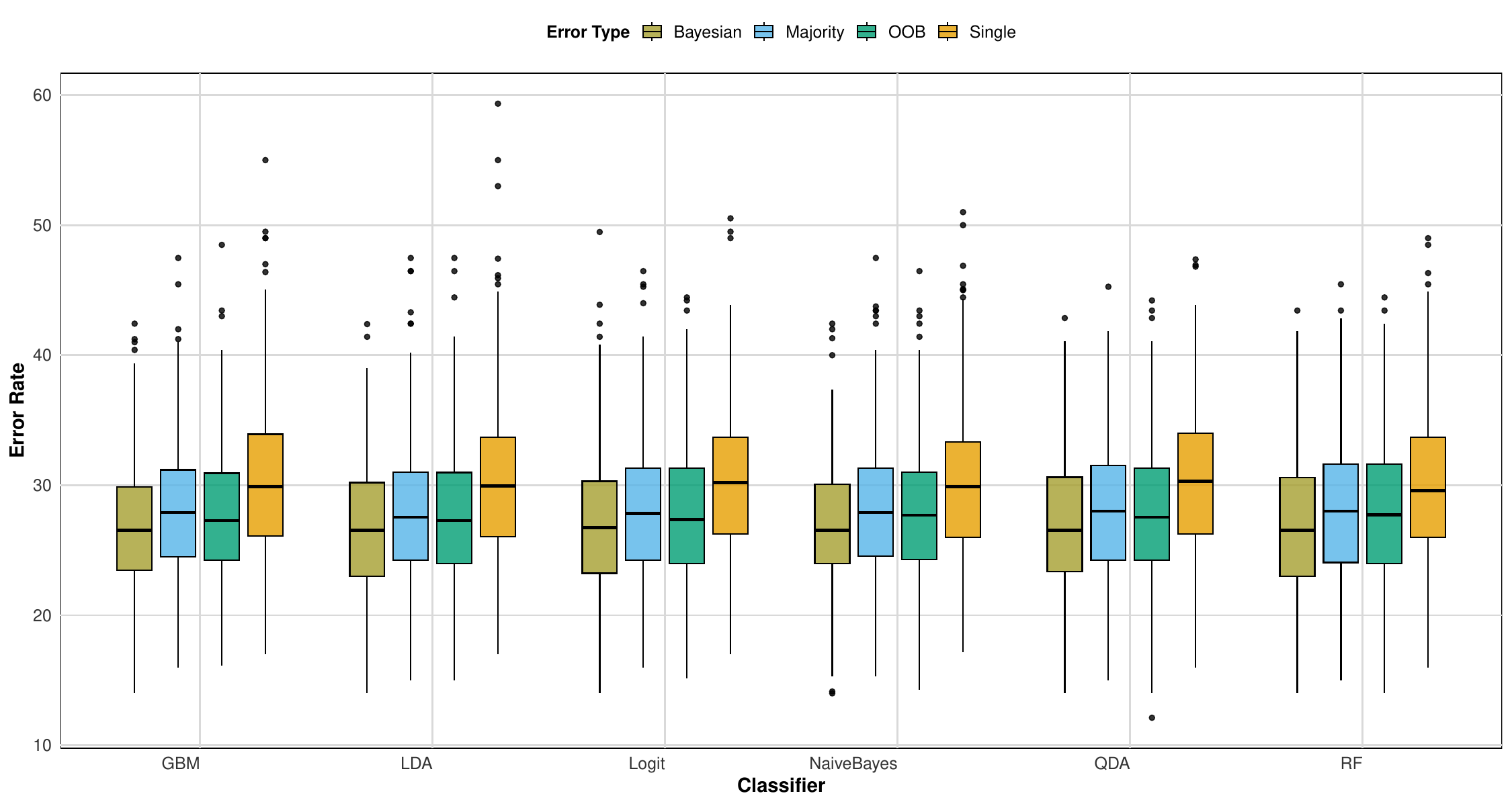}
    \caption{Boxplot of error rates for different classifiers (Logit, LDA, QDA, Naïve Bayes, RF, GBM) across four approaches (single, majority vote, OOB weight, Bayesian) for scenario 6 setting. Each boxplot represents the distribution of error rates computed from 500 replicated simulations.}
    \label{fig:fig6}
\end{figure}

\begin{figure}[H]
    \centering
    \includegraphics[width=0.80\textwidth]{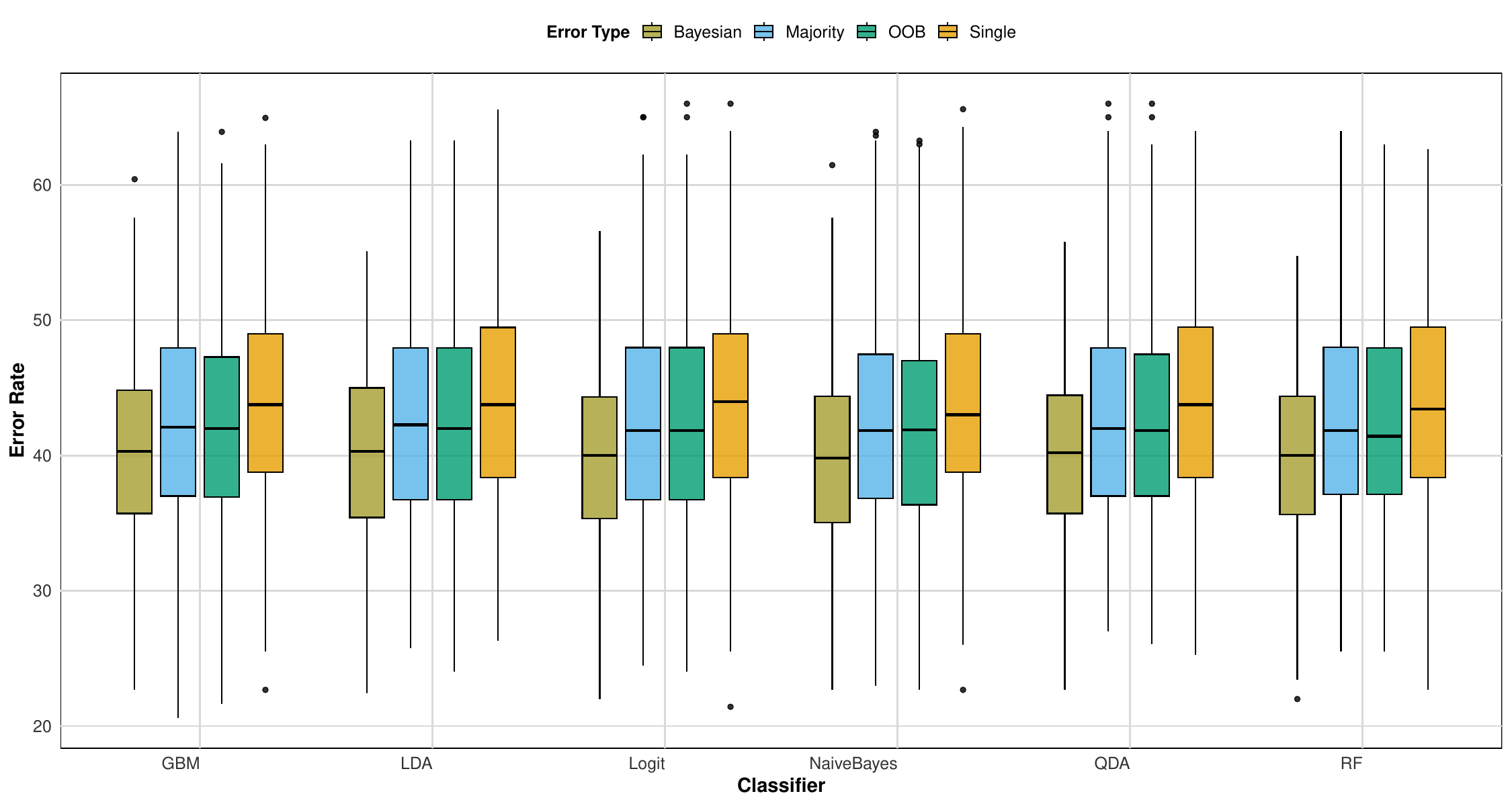}
    \caption{Boxplot of error rates for different classifiers (Logit, LDA, QDA, Naïve Bayes, RF, GBM) across four approaches (single, majority vote, OOB weight, Bayesian) for scenario 7 setting. Each boxplot represents the distribution of error rates computed from 500 replicated simulations.}
    \label{fig:fig7}
\end{figure}

\begin{figure}[H]
    \centering
    \includegraphics[width=0.80\textwidth]{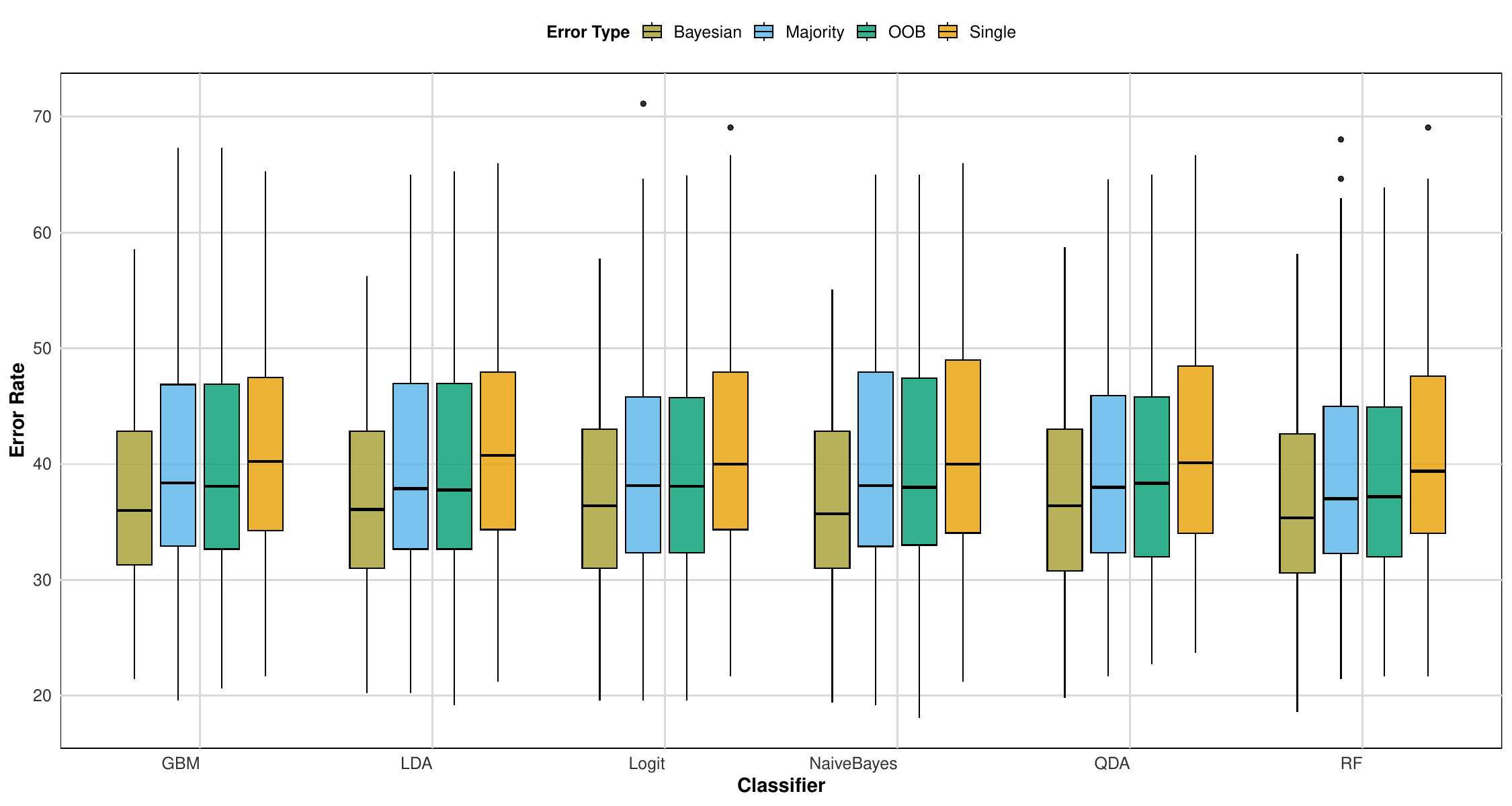}
    \caption{Boxplot of error rates for different classifiers (Logit, LDA, QDA, Naïve Bayes, RF, GBM) across four approaches (single, majority vote, OOB weight, Bayesian) for scenario 8 setting. Each boxplot represents the distribution of error rates computed from 500 replicated simulations.}
    \label{fig:fig8}
\end{figure}

\begin{figure}[H]
    \centering
\includegraphics[width=0.80\textwidth]{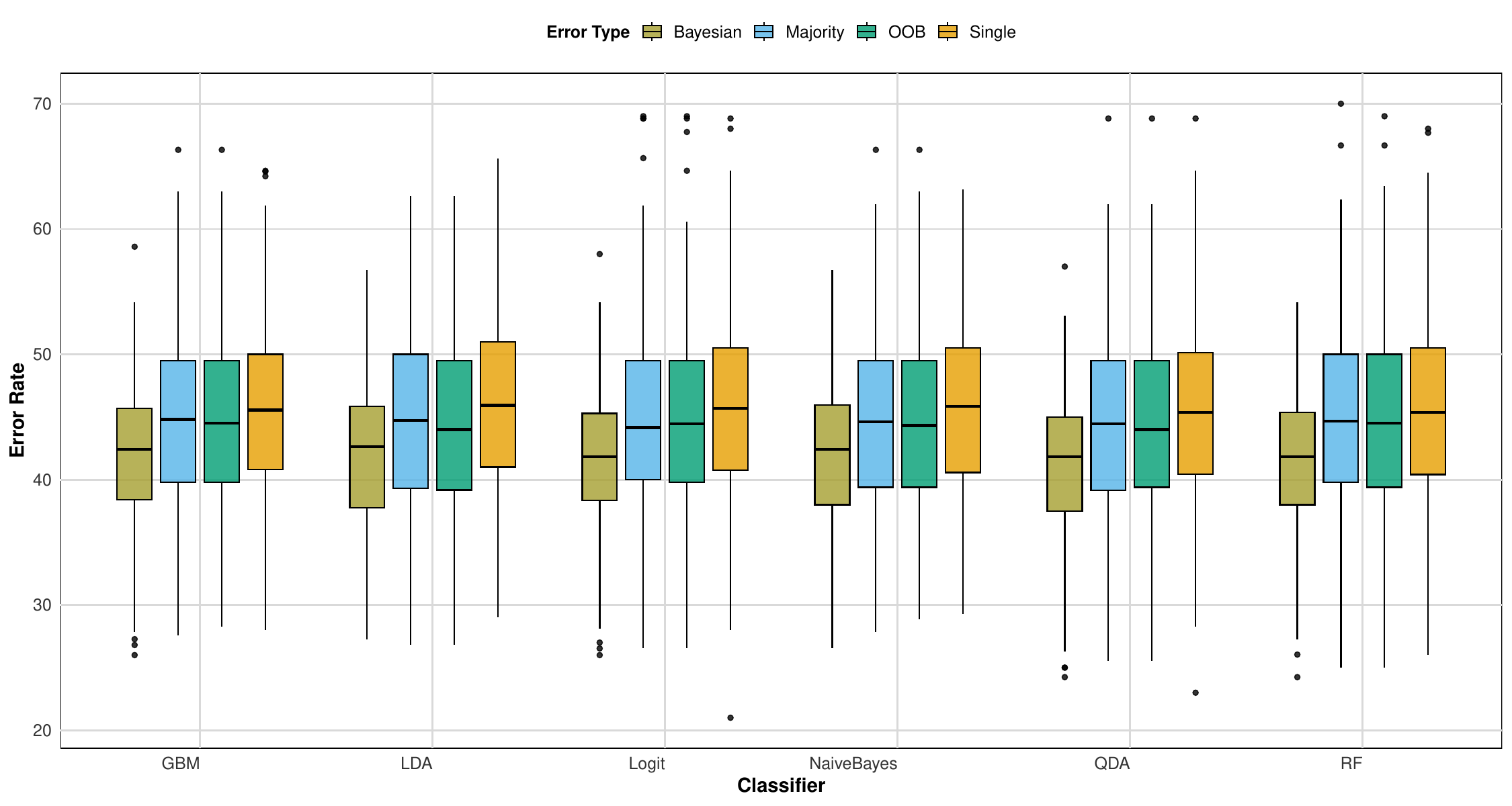}
    \caption{Boxplot of error rates for different classifiers (Logit, LDA, QDA, Naïve Bayes, RF, GBM) across four approaches (single, majority vote, OOB weight, Bayesian) for scenario 9 setting. Each boxplot represents the distribution of error rates computed from 500 replicated simulations.}
    \label{fig:fig9}
\end{figure}
\end{appendices}

\nocite{*} 
\bibliographystyle{unsrt}  
\bibliography{references}  

\end{document}